\documentclass[10pt,a4paper,twocolumn]{article}
\newif\iftwocol \twocoltrue
\iftwocol
  \usepackage[margin=1.9cm]{geometry}
\else
  \usepackage[margin=2.5cm]{geometry}
\fi
\usepackage{amsmath}
\usepackage{newtxtext,newtxmath}
\usepackage[scaled=0.88]{inconsolata}
\usepackage{booktabs}
\usepackage{multirow}
\usepackage{array}
\usepackage{xcolor}
\usepackage{microtype}
\usepackage{enumitem}
\usepackage{tikz}
\usepackage{pgfplots}
\pgfplotsset{compat=1.18}
\usetikzlibrary{arrows.meta,positioning,fit,backgrounds,shapes.geometric,calc}
\usepackage[font=small,labelfont=bf,skip=6pt]{caption}
\usepackage{subcaption}
\usepackage{abstract}
\usepackage{titlesec}
\usepackage[hidelinks]{hyperref}
\usepackage[capitalise,nameinlink]{cleveref}
\usepackage[numbers,sort&compress]{natbib}
\usepackage{fancyhdr}

\definecolor{sqlcol}{HTML}{6929C4}
\definecolor{veccol}{HTML}{1192E8}
\definecolor{graphcol}{HTML}{0F62FE}
\definecolor{okcol}{HTML}{198038}
\definecolor{warncol}{HTML}{8E6A00}
\definecolor{badcol}{HTML}{DA1E28}
\definecolor{rulecol}{HTML}{525252}

\newcommand{\gpass}{\textcolor{okcol}{$\blacksquare$}}
\newcommand{\ghall}{\textcolor{badcol}{$\blacksquare$}}
\newcommand{\gunc}{\textcolor{warncol}{$\blacksquare$}}
\newcommand{\gref}{\textcolor{gray!65}{$\square$}}
\newcommand{\gfail}{\textcolor{gray!75}{$\times$}}

\titleformat{\section}{\normalfont\Large\bfseries}{\thesection}{0.7em}{}
\titleformat{\subsection}{\normalfont\large\bfseries}{\thesubsection}{0.6em}{}
\titlespacing*{\section}{0pt}{1.6ex plus 0.4ex minus 0.2ex}{0.9ex plus 0.2ex}
\setlist[itemize]{leftmargin=1.35em,itemsep=0.22em,topsep=0.3em}
\setlist[enumerate]{leftmargin=1.5em,itemsep=0.22em,topsep=0.3em}

\iftwocol
  \newenvironment{widefig}{\begin{figure*}}{\end{figure*}}
  \newenvironment{widetab}{\begin{table*}}{\end{table*}}
\else
  \newenvironment{widefig}{\begin{figure}}{\end{figure}}
  \newenvironment{widetab}{\begin{table}}{\end{table}}
\fi

\iftwocol
\fi

\newlength{\abstractwidth}

\begin{document}
\twocolumn[
\begin{center}
{\LARGE\bfseries Beyond Similarity Search:\par}
\vspace{0.35em}
{\LARGE\bfseries Knowledge Graph Metadata Enrichment for Enterprise Retrieval-Augmented Generation\par}
\vspace{1.1em}
{\large Fakabbir Amin\par}
\vspace{0.35em}
{\normalsize\texttt{fakabbir@gmail.com}\par}
\vspace{0.35em}
{\small September 2026\par}
\end{center}
\vspace{0.6em}

\begin{center}
\begin{minipage}{\abstractwidth}
\centerline{\large\bfseries Abstract}
\vspace{0.5em}
\small
A key challenging aspect of RAG systems is to retrieve information from enterprise documents created by multiple teams over a period of time, where it is not possible to enrich the information using the latest information and metadata available outside the scope of organizational documentation responsibility. We investigated whether using a knowledge graph to store metadata and frequently changing information would improve upon knowledge retrieval over similarity search over embeddings. To perform the experiment, we created a snapshot of the SEC EDGAR \citep{sec2024edgar} dataset for a quarter, comprising 167{,}050 filings from 10{,}511 companies and 508{,}714 embedded chunks of risk-factor text. To simulate the enrichment, we created a knowledge graph of 430{,}257 nodes and 561{,}144 edges carrying an accession number on every edge. In order to evaluate the gain obtained via a graph-based design approach, we evaluated three architectures: text-to-SQL over typed columns, dense-vector RAG with HNSW cosine search \citep{lewis2020rag,malkov2018hnsw}, and a two-stage GraphRAG design. All architectures share the same model and temperature to allow comparison. The results obtained are scored programmatically by an LLM as judge over 20 questions divided across four different categories, each question asked three times. The GraphRAG-based architecture solved 16 of 20 questions on all three attempts, compared to 10 for text-to-SQL and 3 for vector RAG. GraphRAG lost on structured aggregation queries. Vector RAG lost the single-document category it was expected to win, passing 3 of 15 where GraphRAG passed 15 of 15, because ``belongs to company $X$'' is not recoverable from cosine similarity at this scale. All six confident falsehoods in the benchmark were text-to-SQL conflating distinct people who share a surname, the failure that resolving identity to an identifier before reading any text prevents structurally; GraphRAG made none. We claim no advantage beyond this corpus, this model and this question set.

\vspace{0.9em}

\end{minipage}
\end{center}
\vspace{1.0em}
]
\section{Introduction}
\label{sec:intro}

A widely followed process to build context for a query is to use retrieval-augmented
generation to extract information and then provide it to a language model to construct
a result \citep{lewis2020rag,gao2023ragsurvey}, including recent iterative and corrective
retrieval paradigms \citep{asai2024selfrag,jiang2023active,yan2024corrective}. The standard
procedure is to encode passages into vectors using an embedding model and then match
the query embedding vector to retrieve the nearest neighbours. It works well if the
returned list of passages contains the answer, provided the model is able to extract
the result from it.

A large number of enterprise document stores face a different problem in terms of
data policy and lifecycle. This includes strict requirements on what to store and when
to store it, making the process of retrieval using an LLM on standard passage collections
inefficient. To simulate this scenario where multiple categories of data coexist,
regulatory disclosures and financial corpora such as SEC EDGAR have increasingly
become demanding benchmarks for factual reasoning and QA \citep{loukas2021edgar,chen2021finqa,islam2023financebench}.

In a real-world scenario, a financial analyst may actually ask: \emph{for every company
where a particular person is an officer or director, list that company's disclosed
subsidiaries and summarize the cybersecurity risks it reported.} To answer this question,
three distinct facts are required: (1)~the person's role is disclosed in an insider-ownership
filing; (2)~the subsidiaries are disclosed in an exhibit attached to the annual report; and
(3)~the risks are narrative prose inside a numbered section of that same report.
In order to obtain this information, three separate documents, in three formats,
produced by three different processes are required.

While the passages could be ranked using various retrieval approaches, the difficulty
is that no single passage contains the combination. We measured this directly on our
corpus: the surname of the subject of our hardest question appears in \textbf{0 of 5{,}210}
extracted risk-factor sections, and only \textbf{1 of 151} subsidiary names appears
anywhere in its own parent's risk-factor section. Ownership and risk are, empirically,
disjoint. \Cref{fig:containers} shows the structure that produces this.

The implication is that a better encoder cannot fix it, and neither can a longer context
window \citep{liu2024lost}. Retrieval can only return text that exists. If the conjunction
of facts a question needs is never co-located in any passage, then ranking passages more
accurately retrieves the same insufficient evidence. The ceiling is a property of the
corpus, not of the retriever.

The limitation arises from an inherent property of the text and data sources, not the
retriever. The second failure observed in enterprise settings is that similarity search
has no mechanism for filtering by entity, because ``belongs to company $X$'' is not a
semantic property of a passage.

Looking into the SEC EDGAR data, two risk-factor sections about supply-chain disruption
are near each other in embedding space even if they belong to entirely different companies.
While for smaller datasets this issue might be alleviated by increasing the top-$k$
candidate budget, we show that it bites hard at 508{,}714 chunks drawn from 5{,}210
filers, and---more usefully---that it bites on exactly the questions where similarity
search is supposed to be the right tool.

This problem could theoretically be addressed by embedding extensive metadata into each
chunk and refining the retrieval and ranking process. However, in an enterprise setting,
such metadata is often dynamic in nature, and continuously re-indexing or re-embedding
processed chunks is computationally prohibitive and operationally impractical. An
alternative is to formulate the data storage layer to decouple dynamic metadata and
relational structure from the passage chunks stored in the vector database.

This leads directly to storing the structure explicitly: resolve entities to stable
identifiers, materialize the relationships as edges, and attach a provenance key to
every edge so that any claim can be traced to the document that supports it. This is
the family of designs now loosely called GraphRAG \citep{edge2024graphrag,peng2024graphragsurvey}
and knowledge graph-augmented generation \citep{pan2024unifying,gutierrez2024hipporag,he2024gretriever}.
The question this paper addresses is not whether that is possible but whether it is
\emph{worth it}, and on which question shapes---measured against two baselines that were
given every reasonable advantage.

We present an empirical systems study on one corpus with one model and twenty questions.
It contributes measurements and diagnosed failure modes, not a new algorithm. Our
contributions are:

\begin{itemize}
  \item \textbf{A corpus and benchmark with hard entity identity.} Twelve monthly
        EDGAR feeds ingested into a relational store and a graph
        (\Cref{tab:corpus}), with 20 questions in four categories. The multi-hop
        subjects were selected on measured evidence: the surname must appear in
        \emph{zero} risk-factor sections and must be shared by two or more
        distinct filer identifiers, so that resolving identity by name is
        demonstrably wrong rather than merely inelegant
        (\Cref{sec:design}).
  \item \textbf{A matched three-way comparison.} 180 programmatically scored
        attempts under one model at temperature 0, with the fan-out guardrail and
        repair policy applied identically to both query-writing arms
        (\Cref{sec:arch}), and with two of the four categories chosen so that the
        graph should lose.
  \item \textbf{A result that contradicts our expectation, reported with the
        same prominence as the wins.} Vector RAG lost the single-document
        category it was expected to win, passing 3 of 15 attempts where GraphRAG
        passed 15 of 15 (\Cref{sec:failures}).
  \item \textbf{A cost accounting.} GraphRAG's accuracy is bought:
        520{,}929 tokens against text-to-SQL's
        144{,}068, and a repair call triggered on
        19 of 60 attempts against 5 of 60
        (\Cref{tab:main}, \Cref{tab:repair}). We also report the reverse cost:
        text-to-SQL's generated joins reached 274\,s on a
        single attempt.
  \item \textbf{A safety result that does not follow the accuracy result.}
        All six confident falsehoods were text-to-SQL, all on the multi-hop
        category, and all the same error: conflating distinct people who share a
        surname. GraphRAG made none. Resolving identity to a stable identifier
        before reading any text prevents that class of error structurally rather
        than probabilistically (\Cref{sec:failures}).
\end{itemize}

\Cref{sec:related} places the work; \Cref{sec:dataset} and \Cref{sec:arch}
describe the corpus and the three architectures; \Cref{sec:design} gives the
protocol; \Cref{sec:results} reports what happened; \Cref{sec:failures} explains
the failures; \Cref{sec:threats} states what these measurements cannot support.

\begin{widefig}[htbp]
\centering
\begin{tikzpicture}[
  font=\footnotesize,
  >={Stealth[length=1.7mm,width=1.3mm]},
  doc/.style={draw=gray!55, rounded corners=1.5pt, fill=white, align=center,
              font=\scriptsize, inner sep=3pt, text width=3.9cm, minimum height=1.0cm},
  emb/.style={doc, draw=veccol!70, fill=veccol!10, line width=0.7pt},
  noemb/.style={doc, draw=warncol!75, fill=warncol!10, line width=0.7pt},
  sep/.style={doc, draw=sqlcol!70, fill=sqlcol!10, line width=0.7pt},
  wrap/.style={draw=gray!45, dashed, rounded corners=3pt, inner sep=6pt},
  edge/.style={draw=graphcol!75, rounded corners=1.5pt, fill=graphcol!12,
               align=center, font=\tiny, inner sep=2.5pt, text width=3.3cm,
               minimum height=0.66cm, line width=0.6pt},
  ann/.style={align=center, font=\tiny, text width=5.2cm, inner sep=1pt},
  arr/.style={->, draw=gray!60, line width=0.5pt},
]
\node[emb]   (i1a) at (0,0) {\textbf{Item 1A} Risk Factors\\[1pt]
                             \tiny narrative prose, mean 116{,}779 chars};
\node[ann, below=1.5mm of i1a] (a1)
  {\textcolor{veccol}{\textbf{embedded}} $\rightarrow$ 508{,}714 chunks\\
   \textcolor{gray!70}{the whole universe similarity search can reach}};

\node[noemb, below=3.0mm of a1] (ex21) {\textbf{EX-21} Subsidiaries\\[1pt]
                             \tiny structured list: name + jurisdiction};
\node[ann, below=1.5mm of ex21] (a2)
  {\textcolor{warncol}{\textbf{not embedded}} $\rightarrow$ 212{,}124 rows\\
   \textcolor{gray!70}{a list does not benefit from an encoder}};

\node[wrap, fit=(i1a)(a1)(ex21)(a2),
      label={[font=\scriptsize\bfseries, text=gray!80, inner sep=2pt]above:%
      10-K annual report \textmd{\tiny(one accession number)}}] (tenk) {};

\node[sep, below=7mm of tenk.south] (f345) {\textbf{Form 3 / 4 / 5}\\[1pt]
                             \tiny a different filing entirely};
\node[ann, below=1.5mm of f345] (a3)
  {\textcolor{sqlcol}{\textbf{not embedded}} $\rightarrow$ 150{,}917 rows\\
   \textcolor{gray!70}{carries the CIK that separates people sharing a surname}};

\node[edge] (e1) at (7.6,0.0)   {\texttt{:FILED}\\\tiny 167{,}050};
\node[edge] (e2) at (7.6,-1.5)  {\texttt{:SUBSIDIARY\_OF}\\\tiny 212{,}124};
\node[edge] (e3) at (7.6,-3.0)  {\texttt{:OFFICER\_OF} / \texttt{:DIRECTOR\_OF}\\\tiny 86{,}180 / 70{,}167};
\node[wrap, draw=graphcol!55, fit=(e1)(e2)(e3),
      label={[font=\scriptsize\bfseries, text=graphcol, inner sep=2pt]above:%
      graph projection \textmd{\tiny(accession\_number on every edge)}}] (g) {};

\draw[arr] (i1a.east)  -- ++(0.5,0) |- (e1.west);
\draw[arr] (ex21.east) -- ++(0.5,0) |- (e2.west);
\draw[arr] (f345.east) -- ++(0.5,0) |- (e3.west);

\node[draw=badcol!70, fill=badcol!8, line width=0.7pt, rounded corners=2pt,
      align=left, font=\scriptsize, inner sep=5pt, text width=12.4cm,
      anchor=north west] at ($(tenk.north west)+(0,-7.55)$)
  {\textbf{Why no single chunk answers a multi-hop question.} Measured on this
   corpus: the subject's surname appears in \textbf{0 of 5{,}210} Item~1A
   sections, and only \textbf{1 of 151} subsidiary names appears anywhere in its
   own parent's Item~1A. Role, ownership and risk are written in three containers
   by three teams, so the combination exists in no retrievable passage.};
\end{tikzpicture}
\caption{The three facts a multi-hop question needs live in three different
containers. Item~1A is narrative prose and is the only section embedded; EX-21 is
a structured list attached to the same filing; insider roles are in a separate
filing type entirely. Every projected edge carries the accession number of the
filing that asserts it.}
\label{fig:containers}
\end{widefig}

\section{Related Work}
\label{sec:related}

\textbf{Retrieval-augmented generation.} Grounding a generator in retrieved
passages was formalized by \citet{lewis2020rag} and rests on dense retrieval
\citep{karpukhin2020dpr} and sentence-level encoders \citep{reimers2019sbert}.
The literature since is extensive \citep{gao2023ragsurvey}, and much of it optimizes
the retrieval and ranking steps: better encoders, rerankers, hybrid sparse-dense scoring
\citep{robertson2009bm25}, iterative and active retrieval
\citep{asai2024selfrag,jiang2023active}, and corrective retrieval
\citep{yan2024corrective}. Furthermore, language models face documented context-dilution
effects when evidence is scattered across long or multiple retrieved passages
\citep{liu2024lost}. Our negative result on similarity search is orthogonal to that line of
work. We do not claim the retriever ranked badly; we claim that on questions whose
answer spans documents, no ranking of passages can return evidence that no passage
contains, and that entity membership is not a property the embedding encodes.

\textbf{Structured querying as a baseline.} Translating questions into executable
queries has its own long line, from \citet{zhong2017seq2sql} to the cross-domain
Spider benchmark \citep{yu2018spider}, and more recently to large-scale realistic
enterprise database benchmarks such as BIRD \citep{li2023bird} and multi-step
decomposition and self-correction frameworks like DIN-SQL \citep{pourreza2023dinsql}.
In realistic schemas, schema linking, nested joins, and fan-out aggregations present
severe failure modes for LLMs. We treat text-to-SQL not as a competitor to be
defeated but as the strongest available baseline, and we give it the complete schema
including the tables that make the multi-hop answers reachable by joins.
This matters for the interpretation of our results: text-to-SQL wins one of the
four categories outright and beats our graph implementation in a second.

\textbf{Knowledge graphs and graph-based retrieval.} Knowledge graphs
\citep{hogan2021kg} and their synergy with language models \citep{pan2024unifying}
predate current interest, and ``GraphRAG'' is now an overloaded label
\citep{peng2024graphragsurvey}. Recent systems combine graph structure with LLMs
via graph neural reasoning \citep{yasunaga2021qagnn,zhang2022greaselm}, textual
subgraph retrieval \citep{he2024gretriever}, or neurobiologically inspired
associative memory indexing such as HippoRAG \citep{gutierrez2024hipporag}.
The best-known formulation \citep{edge2024graphrag} builds a community structure
over an LLM-extracted graph and is aimed at query-focused summarization --- global
questions such as ``what are the themes in this corpus''. Our design targets a
different shape. The graph is built from filing metadata and structured exhibits
rather than by LLM extraction; the traversal resolves \emph{which entities} are in
scope, and the answer text is then fetched by primary key from the relational store
rather than summarized from the graph. Only the first hop is LLM-authored; expansion is
deterministic, for reasons we quantify in \Cref{sec:failures}.

\textbf{Multi-hop question answering and entity linking.} HotpotQA
\citep{yang2018hotpotqa}, 2WikiMultihopQA \citep{ho2020hotpotqa}, and MuSiQue \citep{trivedi2022musique}
established multi-hop evaluation, generally over Wikipedia, where the hops are between
topically related articles and entity names are largely unambiguous. Similarly,
financial-domain benchmarks such as EDGAR-CORPUS \citep{loukas2021edgar}, FinQA
\citep{chen2021finqa}, TAT-QA \citep{zhu2021tatqa}, and FinanceBench
\citep{islam2023financebench} evaluate numerical and hybrid reasoning over SEC filings.
However, our corpus differs in the property that drives our main result: entity
\emph{identity} is genuinely ambiguous. In dense entity retrieval and zero-shot entity
linking \citep{wu2020blink}, distinguishing mentions without authoritative identifiers
remains difficult. Three distinct people in this corpus share the surname of our
featured subject, and only a stable identifier separates them. A benchmark where names
resolve cleanly cannot expose the failure mode we observe.

\textbf{Evaluation and citation attribution.} Automated RAG evaluation increasingly
uses an LLM judge \citep{es2024ragas}. We deliberately do not, because the object
under study is whether a system states falsehoods, and hallucination in the judge
\citep{ji2023hallucination} would be indistinguishable from hallucination in the
system. Our scoring is a deterministic function of required entities, forbidden
entities, keyword presence and citation correctness (\Cref{sec:design}). This aligns
with recent literature on citation attribution and grounded generation, such as the
ALCE benchmark \citep{gao2023alce} and attribution measurement frameworks
\citep{rashkin2023attribution}, which stress verifiable provenance over qualitative
evaluations. This buys reproducibility at the cost of rewarding only the answers
we anticipated.

Finally, we position this paper honestly: it is a single-corpus empirical study
whose value is in its measurements and its diagnosed failures, not in a new
retrieval algorithm.

\section{Corpus Construction}
\label{sec:dataset}

\subsection{Source and document structure}

The corpus is drawn from the U.S. Securities and Exchange Commission's EDGAR
system \citep{sec2024edgar}, using the structured SEC EDGAR filings dataset
\citep{amin2026secedgar}. While prior NLP research has examined EDGAR text collections
\citep{loukas2021edgar} or targeted numerical question answering over isolated financial
tables and earnings reports \citep{chen2021finqa,zhu2021tatqa,islam2023financebench},
enterprise inquiry typically demands cross-document reasoning across heterogeneous
filings. We ingested twelve monthly XBRL feeds labelled 2025-09 through 2026-08.
Filing dates in the resulting corpus span 2025-08 to 2026-08, because a monthly feed
includes filings dated at the end of the preceding month; \Cref{fig:months} shows
the distribution, with the small 2025-08 bucket being that spillover.

The structural point that motivates this paper is that an annual report is not one
document. A Form 10-K is a container: a mandated sequence of numbered
\emph{Items}, plus attached \emph{Exhibits}, filed under a single accession
number. Three containers matter here.

\textbf{Item 1A, Risk Factors}, is narrative prose in which a company enumerates
what could go wrong. It is the longest narrative section, averaging 116{,}779
characters and reaching 1{,}060{,}461 in this corpus (\Cref{tab:items}). It is the
only section we embed.

\textbf{Exhibit 21} is an attachment, not a section, required by Item 601(b)(21)
of Regulation S-K. It lists subsidiaries and their jurisdiction of incorporation,
so it parses into rows rather than prose. Two properties matter. First, the
distribution is extremely skewed: 275 filers disclose a single subsidiary while
62 disclose more than 500, and the largest discloses 2{,}780
(\Cref{fig:subdist}). Second, and more consequentially,
\emph{EX-21 is incomplete by design}: rule 601(b)(21)(ii) permits a filer to omit
any subsidiary that is not a ``significant subsidiary''. The exhibit is therefore
a floor on corporate structure and never a census, and the graph inherits that
limit. An absent subsidiary is not evidence that none exists.

\textbf{Forms 3, 4 and 5} are separate filings, not part of the 10-K. Officers,
direct directors and ten-percent owners report their positions and transactions in
structured XML. This is the container that makes identity resolvable, because each
filing carries the insider's own Central Index Key (CIK) --- a stable SEC
identifier. Two directors who share a surname are separable on CIK even though
their names are not.

\Cref{fig:containers} summarizes the arrangement and the measured disjointness
that follows from it.

\subsection{Ingestion}

\begin{figure}[htbp]
\centering
\begin{tikzpicture}
\begin{axis}[
  xbar, bar width=7pt, width=0.74\linewidth, height=6.0cm,
  symbolic y coords={{HNSW + graph},{embed 508,714 chunks},{load RDS},{upload to S3},{extract HTML},{Form 3/4/5},{fetch 10,592 docs},{parse 12 feeds},{fetch feeds}}, ytick=data,
  xmin=0, xmax=132, xlabel={Wall-clock minutes}, xlabel near ticks,
  axis lines*=left, tick style={draw=none}, xmajorgrids, grid style={gray!20},
  nodes near coords, nodes near coords style={font=\tiny}, enlarge y limits=0.09,
  y tick label style={font=\footnotesize},
]
\addplot[fill=graphcol] coordinates {(4.00,{HNSW + graph}) (87.00,{embed 508,714 chunks}) (6.00,{load RDS}) (12.00,{upload to S3}) (19.00,{extract HTML}) (0.40,{Form 3/4/5}) (118.00,{fetch 10,592 docs}) (4.20,{parse 12 feeds}) (0.90,{fetch feeds})};
\end{axis}
\end{tikzpicture}
\caption{Ingestion wall-clock, about 252 minutes total. Document fetch and
embedding are roughly 82\% of it. The fetch component is bounded externally by
EDGAR's ten-requests-per-second fair-access policy and cannot be parallelized
away.}
\label{fig:pipeline}
\end{figure}
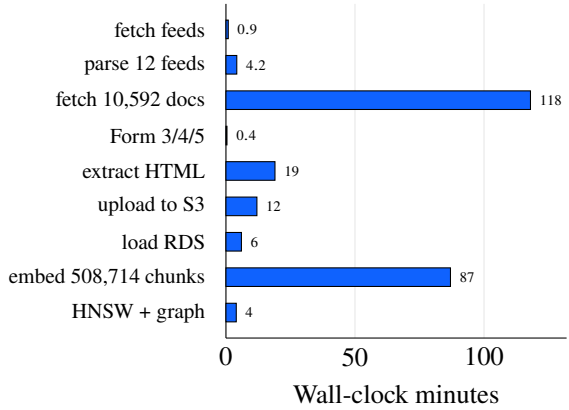

The pipeline runs in nine stages: fetch and parse the monthly feeds; fetch the
documents each filing manifests; extract narrative Items from HTML; parse EX-21
and the ownership forms; stage to object storage; load the relational store;
embed; and build indexes and the graph projection. \Cref{fig:pipeline} gives the
wall-clock breakdown, totalling roughly 252 minutes. Two stages dominate:
document fetch (118 minutes) and embedding (87 minutes), together about 82\% of
the total. The fetch cost is externally bounded --- EDGAR's fair-access policy
caps automated requests at ten per second per requester
\citep{sec2024edgar} --- so no amount of parallelism removes it.

The filings manifest 1{,}904{,}128 documents in total. We fetched and parsed the
10{,}592 that carry the narrative and exhibit content the benchmark needs,
amounting to 19.7\,GB of raw HTML, which reduces to a 3{,}803\,MiB relational
store plus a 1.0\,GB vector index. \Cref{tab:corpus} gives the resulting counts.

\subsection{Schema}

The relational store is PostgreSQL. Filings and companies are keyed by accession
number and CIK. \texttt{filing\_section} holds one row per extracted Item with its
text; \texttt{section\_chunk} holds the embedded chunks; \texttt{subsidiary} holds
parsed EX-21 rows; \texttt{reporting\_owner} holds parsed ownership rows; and
\texttt{filing\_auditor} holds auditor attestations.

The graph is a projection of the same data, not a second source of truth. It holds
430{,}257 nodes across five labels --- \texttt{Filing} (167{,}050),
\texttt{Subsidiary} (200{,}869), \texttt{Person} (51{,}558), \texttt{Company}
(10{,}511) and \texttt{AuditFirm} (269) --- and 561{,}144 edges across seven
types: \texttt{:SUBSIDIARY\_OF} (212{,}124), \texttt{:FILED} (167{,}050),
\texttt{:OFFICER\_OF} (86{,}180), \texttt{:DIRECTOR\_OF} (70{,}167),
\texttt{:OWNS\_SHARES} (19{,}542), \texttt{:AUDITED\_BY} (5{,}835) and
\texttt{:RESOLVES\_TO} (246).

Two design decisions in that schema carry most of the paper's weight.

\emph{Every edge carries the accession number of the filing that asserts it.}
Provenance is a property of the relationship rather than of a document, which is
what lets an answer cite the exact filing supporting each individual claim rather
than the set of documents that were retrieved.

\emph{\texttt{:RESOLVES\_TO} is the entity-resolution layer, and it is thin.}
It links a subsidiary named in an EX-21 to a company that files in its own right,
a fact no single document states. There are 246 such edges against 200{,}869
subsidiary nodes --- a coverage of 0.12\% --- because EX-21 carries no CIK for a
subsidiary, so the link must be established by name matching. This mirrors the classic
challenge of entity disambiguation and cross-document entity linking
\citep{wu2020blink}, where surface names lack authoritative keys. We show in
\Cref{sec:failures} that this thinness is a direct cause of the graph losing the
one category built to exercise it.

\begin{table}[htbp]
\centering
\begin{tabular}{@{}lr@{}}
\toprule
\textbf{Entity} & \textbf{Count} \\
\midrule
Filings (all form types)          & 167,050 \\
Distinct filers (companies)       & 10,511 \\
Documents in filing manifests     & 1,904,128 \\
Documents fetched and parsed      & 10,592 \\
10-K narrative sections extracted & 23,649 \\
Item~1A chunks embedded           & 508,714 \\
EX-21 subsidiary rows             & 212,124 \\
Form 3/4/5 ownership rows         & 150,917 \\
Distinct insiders                 & 51,558 \\
Auditor attestations              & 5,835 \\
\midrule
Graph nodes                       & 430,257 \\
Graph edges                       & 561,144 \\
\midrule
Relational store on disk          & 3,803\,MiB \\
HNSW index                        & 1.0\,GB \\
\bottomrule
\end{tabular}

\caption{The loaded corpus. Twelve monthly EDGAR feeds, 2025-09 through 2026-08.
Node and edge counts are live counts from the graph; the per-type edge breakdown
in \Cref{sec:dataset} sums exactly to the total shown here.}
\label{tab:corpus}
\end{table}

\begin{widetab}[htbp]
\centering
\small
\begin{tabular}{@{}llrrrr@{}}
\toprule
\textbf{Item} & \textbf{Section} & \textbf{Filings} & \textbf{Total chars} & \textbf{Mean} & \textbf{Max} \\
\midrule
1 & Business & 5,539 & 352,550,560 & 63,649 & 717,422 \\
\textbf{1A} & Risk Factors & 5,210 & 608,417,255 & 116,779 & 1,060,461 \\
1B & Unresolved Staff Comments & 63 & 2,000,524 & 31,754 & 127,289 \\
2 & Properties & 2,683 & 12,281,341 & 4,577 & 390,319 \\
3 & Legal Proceedings & 1,518 & 5,126,849 & 3,377 & 117,185 \\
7 & Management's Discussion \& Analysis & 5,609 & 294,551,458 & 52,514 & 473,035 \\
7A & Market Risk & 3,027 & 16,460,821 & 5,438 & 276,843 \\
\bottomrule
\end{tabular}

\caption{Narrative sections extracted per 10-K Item. Only Item~1A (bold) is
embedded, giving 508{,}714 chunks. Its mean length exceeds the encoder's
256-word-piece window by roughly two orders of magnitude, which is why chunking is
forced rather than optional.}
\label{tab:items}
\end{widetab}

\begin{widefig}[htbp]
\centering
\begin{subfigure}[t]{0.48\linewidth}
\centering
\begin{tikzpicture}
\begin{axis}[
  ybar, bar width=15pt, width=\linewidth, height=4.4cm,
  symbolic x coords={{1},{2-5},{6-20},{21-100},{101-500},{500+}}, xtick=data,
  ymin=0, ylabel={Filers}, ylabel near ticks,
  xlabel={Subsidiaries disclosed in EX-21}, xlabel near ticks,
  axis lines*=left, tick style={draw=none}, ymajorgrids, grid style={gray!20},
  nodes near coords, nodes near coords style={font=\tiny}, enlarge x limits=0.13,
]
\addplot[fill=graphcol] coordinates {({1},275) ({2-5},950) ({6-20},1261) ({21-100},1079) ({101-500},439) ({500+},62)};
\end{axis}
\end{tikzpicture}
\caption{EX-21 disclosure per filer.}
\label{fig:subdist}
\end{subfigure}\hfill
\begin{subfigure}[t]{0.48\linewidth}
\centering
\begin{tikzpicture}
\begin{axis}[
  ybar, bar width=7pt, width=\linewidth, height=4.2cm,
  symbolic x coords={{25-08},{25-09},{25-10},{25-11},{25-12},{26-01},{26-02},{26-03},{26-04},{26-05},{26-06},{26-07},{26-08}}, xtick=data,
  x tick label style={rotate=45, anchor=east, font=\tiny},
  ymin=0, ylabel={Filings with parsed documents}, ylabel near ticks,
  axis lines*=left, tick style={draw=none}, ymajorgrids, grid style={gray!20},
  enlarge x limits=0.05,
]
\addplot[fill=veccol] coordinates {({25-08},362) ({25-09},6581) ({25-10},8000) ({25-11},11776) ({25-12},11504) ({26-01},11513) ({26-02},15902) ({26-03},17934) ({26-04},16633) ({26-05},19594) ({26-06},15082) ({26-07},14050) ({26-08},18119)};
\end{axis}
\end{tikzpicture}
\caption{Filings per month.}
\label{fig:months}
\end{subfigure}
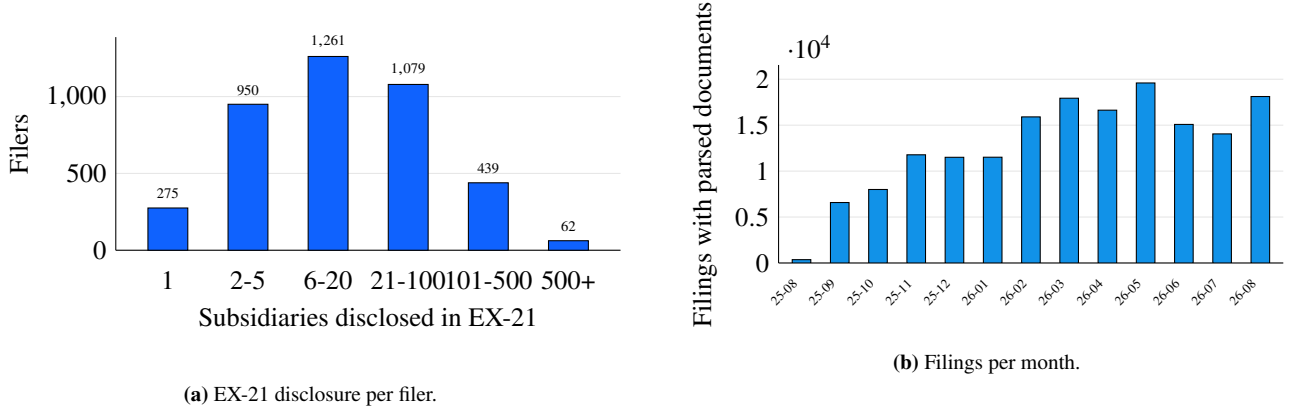
\caption{Corpus shape. (a) EX-21 disclosure is extremely skewed --- 275 filers list one subsidiary, 62 list more than 500 --- so any generated \texttt{LIMIT} over a join to this table starves whole entities unless rows are balanced per entity first. (b) The small 2025-08 bucket is spillover, since a monthly feed contains filings dated at the end of the preceding month.}
\label{fig:corpusdist}
\end{widefig}

\section{System Architecture}
\label{sec:arch}

Three retrieval arms answer the same question over the same stores.
\Cref{fig:arch} shows all three.

\subsection{Text-to-SQL}

The model is shown the complete relational schema, including the
\texttt{subsidiary} and \texttt{reporting\_owner} tables, and writes a SQL query;
the query is executed; the resulting rows are given back to the model, which
writes the answer. Two LLM calls. The schema is a static prompt hardcoded in
the retrieval module, not dynamically introspected from
\texttt{information\_schema} at query time. The query-generation system prompt
specifies rules including: output only SQL, never write data (\texttt{SELECT} /
\texttt{WITH} only), always \texttt{LIMIT} to at most 50 rows, never select raw
\texttt{section\_text} (averaging 75{,}000 characters), prefer boolean flags over
parsing the \texttt{relationship} string, and aggregate per entity to avoid the
fan-out \texttt{LIMIT} trap. The answer-synthesis system prompt instructs the model
to act as a financial-disclosure analyst, answer using only the query result rows,
cite accession numbers, and stay under 300 words. An additional safety guardrail
enforces read-only execution: a regex rejects any statement containing write
keywords (\texttt{INSERT}, \texttt{UPDATE}, \texttt{DELETE}, \texttt{DROP},
\texttt{ALTER}, \texttt{TRUNCATE}, \texttt{CREATE}, \texttt{GRANT},
\texttt{COPY}), and results are fetched up to a 2{,}000-row limit after
\texttt{raise\_limit} raises the model's trailing \texttt{LIMIT}.

This is deliberately a strong baseline rather than a straw man. Because the schema
is complete, every multi-hop answer in the benchmark \emph{is} reachable by joins.
What the model must do unaided is choose the join path, decode a multi-valued
relationship string, and recognize that traversing an ownership chain of unknown
depth requires a recursive CTE. This setup mirrors the realistic schema linking and
multi-table join complexity characterized in database benchmarks such as BIRD
\citep{li2023bird} and decomposition strategies in DIN-SQL \citep{pourreza2023dinsql}.
The schema documentation also states the surname-first convention used for person
names --- an asymmetry in an earlier iteration, where only the graph documentation
mentioned it, is recorded among the harness defects in \Cref{sec:repro}.

\subsection{Vector RAG}

The question is embedded and the $k{=}8$ nearest chunks are retrieved by cosine
similarity from an HNSW index \citep{malkov2018hnsw,pgvector2024}; the chunks are
given to the model, which answers. One LLM call, and no query language, therefore
no repair path. The system prompt instructs the model to act as a
financial-disclosure analyst, answer only from the provided excerpts, cite
accession numbers for every claim, never guess relationships or roles not stated
verbatim in the excerpts, and stay under 300 words. Retrieved chunks are packed
into the context up to a budget of 9{,}000 characters, each formatted with its
accession number, company name, CIK, Item code, chunk index and similarity
score.

The encoder is \texttt{all-MiniLM-L6-v2}, a six-layer 22.7M-parameter model
distilled in the MiniLM family \citep{wang2020minilm} and trained as a sentence
encoder \citep{reimers2019sbert}. It emits 384 dimensions with mean pooling and
$L^2$ normalization, so cosine similarity reduces to a dot product. The model
is loaded locally via \texttt{sentence\_transformers.SentenceTransformer} rather
than called as an API; encoding is serialized with a thread lock and
\texttt{torch.set\_num\_threads(1)} to prevent crashes from concurrent forward
passes in the multi-threaded server.

One configuration detail is load-bearing. The encoder accepts 256 word-pieces,
roughly a thousand characters, while Item 1A averages 116{,}779 characters. The
section therefore cannot be embedded whole: text is chunked at 1{,}400 characters
with 200 characters of overlap, so a single risk-factor section becomes on the
order of a hundred vectors, and 5{,}210 sections become 508{,}714 chunks. The tension
between fixed embedding token limits and long-document retrieval is well documented
in the RAG literature \citep{gao2023ragsurvey,liu2024lost}. We embed only Item 1A.
That is not a handicap imposed on the baseline but the opposite --- it is the only
section any benchmark question asks about, so restricting the index concentrates
the candidate pool and makes the vector arm \emph{stronger} than it would be with
all seven Items embedded.

\subsection{GraphRAG}

Two LLM calls, as with text-to-SQL, but the work between them is split
differently.

\begin{enumerate}
  \item \textbf{Identify.} The model writes a Cypher query whose job is to
        determine \emph{which entities} the question concerns --- typically
        resolving a person to a CIK and finding the companies where they hold a
        role, with their relevant filings.
  \item \textbf{Expand (deterministic).} The system, not the model, expands each
        identified entity: for every company returned, fetch its subsidiaries and
        its risk-factor filings.
  \item \textbf{Fetch text by key.} The accession numbers from the traversal are
        used to select section text by primary key, and windows of $\pm 900$
        characters around matched terms are extracted, at most three per filing.
        Concretely, the snippet extractor performs a case-insensitive regex search
        for each term from \texttt{text\_terms} (2--4 keyword phrases generated by
        the planning LLM call) in the full section text, takes 450 characters
        before and 450 after each match, enforces non-overlapping windows, and
        returns up to three snippets per filing. If no term matches in a filing's
        text, no snippet is returned for that filing.
        This is a lookup, not a search: the graph has already decided which
        documents are in scope, so there is nothing to rank and nothing to miss.
  \item \textbf{Answer.} The model receives the graph facts and the exact
        passages, and writes the answer with citations.
\end{enumerate}

Stage 2 being system-authored rather than LLM-authored is the design's central
claim, and it is a response to a measured failure rather than an aesthetic
preference. An LLM writing a single query over a fan-out join will attach a
\texttt{LIMIT}, and one entity's hundreds of child rows then consume the entire
budget --- after which the model reports, fluently and wrongly, that the other
entities have no subsidiaries or do not exist. Splitting identification from
expansion makes that class of error structurally impossible rather than something
to be prompted against. This separation contrasts with approaches that delegate
multi-hop traversal or subgraph synthesis entirely to generative prompting
\citep{edge2024graphrag,he2024gretriever}, while sharing the motivation of
associative knowledge graph indexing explored in HippoRAG \citep{gutierrez2024hipporag}.

The stage-1 planning prompt instructs the model to return a JSON object
containing one read-only Cypher query and 2--4 keyword phrases for text search.
Key rules enforced by the prompt: return one row per entity with
\texttt{companyCik} (the system expands every entity completely in stage~2), do
not traverse down to children, \texttt{LIMIT 25} or fewer, follow documented
relationship directions exactly, and never join one edge's \texttt{filingId} to
another's.  The stage-2 expansion is a fixed Cypher query (\texttt{EXPAND\_CQL})
that takes the list of CIKs from stage~1 and, for each company, retrieves its
subsidiary count, a sample of up to 12 subsidiary names, its EX-21 filing
identifier, all risk-factor filing accession numbers, up to 6 insiders, and any
\texttt{:RESOLVES\_TO} reverse links.  This expansion is marked as authoritative
in the final prompt so the model does not infer absence from a \texttt{LIMIT}ed
traversal. The answer prompt instructs the model to act as a
financial-disclosure analyst, answer only from the graph facts and verbatim
excerpts, cite accession numbers for every claim, name the traversal path, and
stay under 300 words.

The repair contract for GraphRAG includes an additional trigger beyond exceptions
and zero rows: if the stage-1 query returns rows but none of the returned columns
is a company CIK, the repair fires, because stage~2 cannot expand entities it
cannot identify. This prevented the failure mode where the model returned 2{,}000
rows of names with no identifiers, leading to a confident ``none exist'' answer.

\subsection{Shared guardrails}

The same failure threatens both query languages, so the same two transforms are
applied to both, identically. Any trailing \texttt{LIMIT} in a generated query is
raised to 2{,}000 rows before execution; the result set is then balanced to at
most 6 rows per entity and 48 rows overall, where the entity column is detected by
name. Applying these to SQL and Cypher alike means the comparison measures
retrieval architecture rather than which prompt happened to dodge the trap.

Both query-writing arms also get exactly one repair attempt, triggered when a
query raises an exception or returns zero rows: the model sees the error text and
rewrites once. There is no second repair. \Cref{sec:results} shows the two arms
used this allowance very unequally.

\subsection{Model access}

All three arms call a single OpenAI-compatible endpoint provided by Bifrost, an
open-source LLM gateway, which routes to DeepSeek or to AWS Bedrock. The
benchmarked configuration is \texttt{deepseek-v4-flash} at temperature 0 with
reasoning disabled, 700 output tokens for query generation (800 for
GraphRAG's Cypher planning) and 1{,}500 for answers. The client is
\texttt{langchain\_openai.ChatOpenAI} with a 180-second timeout and up to 2
retries per call. LLM clients are cached by the tuple
\texttt{(temperature, max\_tokens, reasoning)}, so each unique configuration
creates exactly one client instance. Token accounting is per-thread: each
retrieval mode runs in its own thread, and a thread-local counter prevents the
cross-attribution bug that afflicted an earlier version of the harness
(\Cref{sec:harness}).

Routing through one endpoint is a methodological device, not a convenience. It
makes ``all three arms used the same model with the same settings'' a structural
property of the deployment rather than a convention maintained by discipline: the
model is a request parameter read from one place by all three arms. It also makes
the cross-model probe in \Cref{sec:failures} a configuration change rather than a
code change. Reasoning was disabled because an uncapped reasoning trace consumed
roughly 33{,}000 completion tokens per call, which \texttt{max\_tokens} does not
bound; the setting is identical for every arm, so the comparison is unaffected.

\begin{widefig}[htbp]
\centering
\resizebox{\linewidth}{!}{\begin{tikzpicture}[
  font=\footnotesize,
  >={Stealth[length=1.7mm,width=1.3mm]},
  box/.style={draw=gray!55, line width=0.4pt, rounded corners=1.5pt, fill=white,
              align=center, font=\scriptsize, inner sep=3pt,
              text width=3.9cm, minimum width=4.3cm, minimum height=0.95cm},
  slim/.style={box, minimum height=0.6cm, fill=gray!6, draw=gray!45},
  shared/.style={box, text width=11cm, minimum width=11.5cm, minimum height=0.85cm,
                 fill=gray!8, draw=gray!55},
  gapbox/.style={box, dashed, draw=gray!55, fill=gray!4, text=gray,
                 minimum height=2.55cm},
  key/.style={box, line width=1pt, draw=okcol!70!black, fill=okcol!18,
              minimum height=1.15cm},
  tag/.style={draw=okcol!70!black, fill=okcol!22, rounded corners=1pt,
              inner sep=1.5pt, font=\tiny\bfseries},
  dbicon/.style={cylinder, shape border rotate=90, aspect=0.32, draw=gray!60,
                 fill=gray!18, line width=0.35pt, inner sep=0pt,
                 minimum width=3.8mm, minimum height=2.6mm},
  lane/.style={rounded corners=3pt, inner sep=4pt, line width=0.5pt},
  arr/.style={->, draw=gray!65, line width=0.5pt},
  wire/.style={draw=gray!65, line width=0.5pt},
]

\node[shared] (q) at (5.05,0)
  {\textbf{Natural-language question over the 10-K / insider corpus}\\[1pt]
   \tiny identical prompt, identical model and $T{=}0$ for all three arms};

\node[box] (l1a) at (0,-2.15)
  {LLM writes SQL\\[1pt]\tiny complete schema in prompt; one repair retry};
\node[box] (l1b) at (0,-3.50)
  {PostgreSQL 17\\[1pt]\tiny execute; 45\,s statement timeout};
\node[slim] (l1c) at (0,-4.70) {result rows \tiny(LIMIT capped at 2{,}000)};
\node[gapbox] (l1d) at (0,-6.70)
  {\itshape no deterministic stage\\[3pt]
   \tiny\upshape whatever join the model happened to write \emph{is} the retrieval:
   one bad predicate silently returns a well-formed, incomplete row set};
\node[box] (l1e) at (0,-8.85)
  {LLM answers\\[1pt]\tiny from the returned rows only};

\node[box] (l2a) at (5.05,-2.15)
  {MiniLM-L6-v2 encoder\\[1pt]\tiny 384-d, mean-pooled, normalised (no LLM)};
\node[box] (l2b) at (5.05,-3.50)
  {HNSW cosine top-8\\[1pt]\tiny pgvector over 508{,}714 chunks};
\node[slim] (l2c) at (5.05,-4.70) {8 chunks \tiny(1{,}400 chars, 200 overlap)};
\node[gapbox] (l2d) at (5.05,-6.70)
  {\itshape no query language\\[3pt]
   \tiny\upshape nothing to repair and nothing to expand: the eight nearest
   neighbours are the entire evidence set, whatever the question asked for};
\node[box] (l2e) at (5.05,-8.85)
  {LLM answers\\[1pt]\tiny from the retrieved chunks only};

\node[box] (l3a) at (10.1,-2.15)
  {LLM writes stage-1 Cypher\\[1pt]\tiny \emph{identify entities only}; one repair retry};
\node[box] (l3b) at (10.1,-3.50)
  {Neo4j 5.26 + APOC\\[1pt]\tiny match entity nodes};
\node[slim] (l3c) at (10.1,-4.70) {entity nodes \tiny(seeds, no text yet)};
\node[key] (l3d) at (10.1,-6.00)
  {\textbf{Deterministic expansion}\\[1pt]
   \tiny system code, not the LLM: fixed traversal per seed, 6 filings each, 48 total};
\node[box, minimum height=1cm] (l3e) at (10.1,-7.475)
  {Exact text fetch\\[1pt]\tiny by \texttt{accession\_number} primary key from
   PostgreSQL; 900-char window};
\node[box] (l3f) at (10.1,-8.85)
  {LLM answers\\[1pt]\tiny graph facts + exact snippets};

\begin{scope}[on background layer]
  \node[lane, fill=sqlcol!7,   draw=sqlcol!40,   fit=(l1a)(l1b)(l1c)(l1d)(l1e)] (L1) {};
  \node[lane, fill=veccol!7,   draw=veccol!40,   fit=(l2a)(l2b)(l2c)(l2d)(l2e)] (L2) {};
  \node[lane, fill=graphcol!7, draw=graphcol!40, fit=(l3a)(l3b)(l3c)(l3d)(l3e)(l3f)] (L3) {};
\end{scope}

\node[anchor=south west, font=\footnotesize\bfseries, text=sqlcol!75!black]   at (L1.north west) {Text-to-SQL};
\node[anchor=south west, font=\footnotesize\bfseries, text=veccol!75!black]   at (L2.north west) {Vector RAG};
\node[anchor=south west, font=\footnotesize\bfseries, text=graphcol!75!black] at (L3.north west) {GraphRAG};
\node[anchor=south east, font=\tiny, text=gray!65] at (L1.north east) {2 LLM calls};
\node[anchor=south east, font=\tiny, text=gray!65] at (L2.north east) {1 LLM call};
\node[anchor=south east, font=\tiny, text=gray!65] at (L3.north east) {2 LLM calls};

\node[dbicon, anchor=north east] at ([shift={(-2pt,-2pt)}]l1b.north east) {};
\node[dbicon, anchor=north east] at ([shift={(-2pt,-2pt)}]l2b.north east) {};
\node[dbicon, anchor=north east] at ([shift={(-2pt,-2pt)}]l3b.north east) {};
\node[dbicon, anchor=north east] at ([shift={(-2pt,-2pt)}]l3e.north east) {};

\node[tag, anchor=south east] at ([yshift=1.5pt]l3d.north east) {not LLM-written};

\coordinate (b1) at (5.05,-0.95);
\draw[wire] (q.south) -- (b1);
\draw[arr] (b1) -| (l1a.north);
\draw[arr] (b1) -| (l2a.north);
\draw[arr] (b1) -| (l3a.north);

\draw[arr] (l1a) -- (l1b);  \draw[arr] (l1b) -- (l1c);
\draw[arr] (l1c) -- (l1d);  \draw[arr] (l1d) -- (l1e);
\draw[arr] (l2a) -- (l2b);  \draw[arr] (l2b) -- (l2c);
\draw[arr] (l2c) -- (l2d);  \draw[arr] (l2d) -- (l2e);
\draw[arr] (l3a) -- (l3b);  \draw[arr] (l3b) -- (l3c);
\draw[arr] (l3c) -- (l3d);  \draw[arr] (l3d) -- (l3e);
\draw[arr] (l3e) -- (l3f);

\coordinate (b2) at (5.05,-9.80);
\draw[wire] (l1e.south) |- (b2);
\draw[wire] (l2e.south) |- (b2);
\draw[wire] (l3f.south) |- (b2);
\node[shared] (ans) at (5.05,-10.45)
  {\textbf{Answer}\\[1pt]\tiny scored programmatically -- entity recall, required
   keywords, forbidden entities, citation validity (no LLM judge)};
\draw[arr] (b2) -- (ans.north);

\end{tikzpicture}}
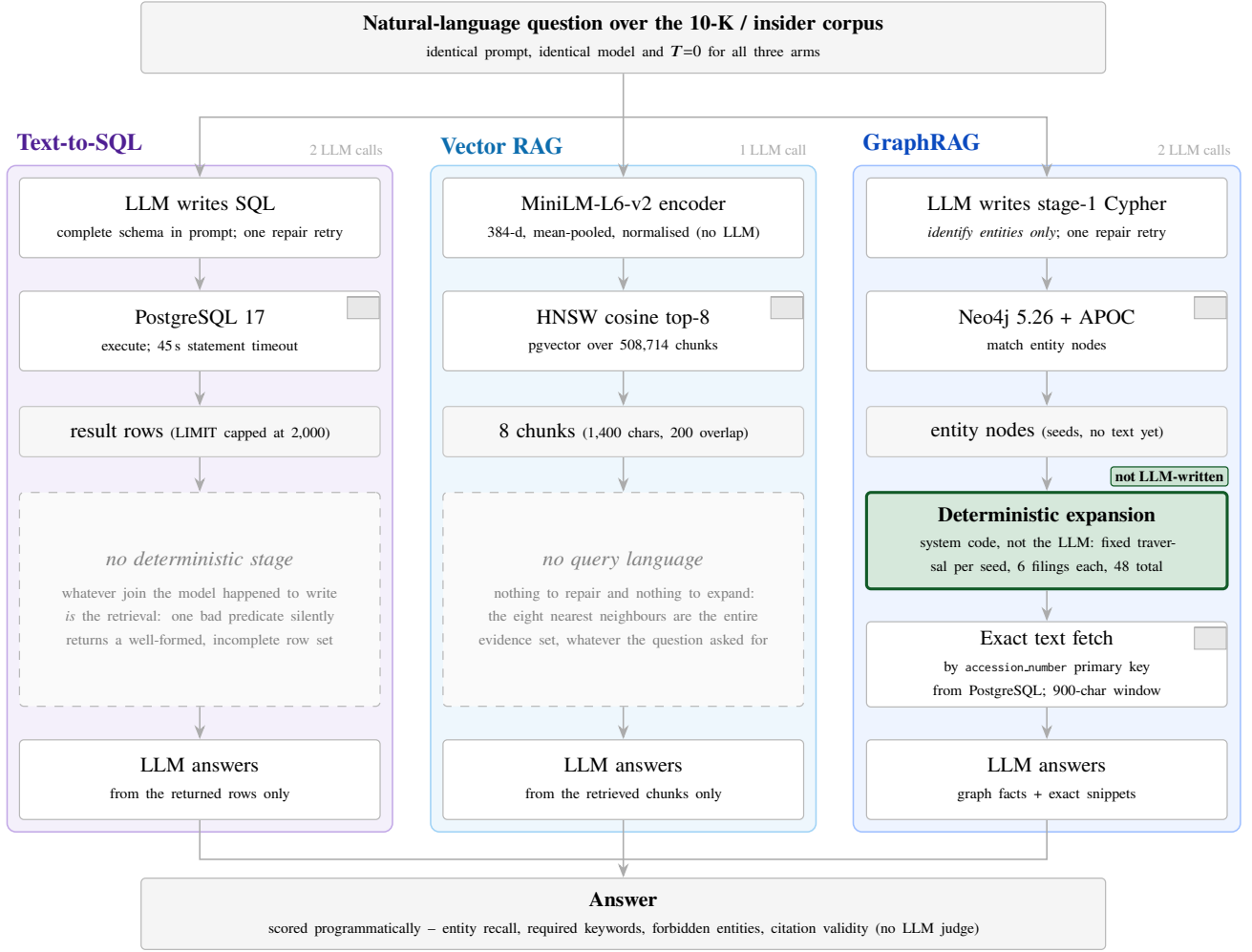
\caption{The three retrieval arms over shared stores. Text-to-SQL and GraphRAG use
two LLM calls each; vector RAG uses one and has no query language to repair. In
GraphRAG only the first hop is LLM-authored: entity expansion is deterministic and
system-authored, which is what prevents one entity's fan-out from consuming the
row budget and the model then reporting that the other entities do not exist.}
\label{fig:arch}
\end{widefig}

\section{Experimental Design}
\label{sec:design}

\subsection{Four question shapes}

The benchmark has 20 questions in four categories of five, listed in full in
\Cref{app:questions}. Each category carries a pre-registered expectation of which
architecture should win, and two of the four were chosen so that the graph is the
\emph{wrong} tool:

\begin{itemize}
  \item \textbf{T1, single-document semantic.} What did one named company disclose
        about one topic in its risk factors? The whole answer lies inside one
        document. Expected winner: vector RAG or GraphRAG.
  \item \textbf{T2, structured aggregation.} Which audit firm serves the most
        companies; which jurisdiction of incorporation is most common. These are
        \texttt{GROUP BY} queries over typed columns. Expected winner:
        text-to-SQL, and a traversal should add nothing.
  \item \textbf{T3, multi-hop relational plus text.} For every company where a
        named person holds an insider role, list its subsidiaries and summarize
        its reported cybersecurity risks. Expected winner: GraphRAG.
  \item \textbf{T4, cross-document entity resolution.} Is a named subsidiary also
        an SEC filer in its own right, and which parent lists it? Expected
        winner: GraphRAG.
\end{itemize}

\subsection{Selecting subjects that make identity hard}

The T3 subjects were not chosen for convenience. Two conditions had to hold
simultaneously, both checked against the full-text index over all 5{,}210
risk-factor sections:

\begin{enumerate}
  \item the surname appears in \emph{zero} risk-factor sections, so no similarity
        search can reach the person at all; and
  \item two or more distinct filer identifiers share that surname, so resolving
        identity by name is demonstrably wrong rather than merely imprecise.
\end{enumerate}

Candidates were rejected on the first condition when the surname was too common in
the text (32, 49, 63 and 109 matching sections respectively) and on the second
when the surname turned out to be unique in the corpus. One unique-surname subject
was retained deliberately as a control: same question shape, no identity trap.

\subsection{Scoring}

Scoring is a deterministic function of the answer text. There is no LLM judge, for
the reason given in \Cref{sec:related}. An attempt \textbf{passes} when all four
conditions hold:

\begin{enumerate}
  \item every required entity is named (entity recall $=1.0$);
  \item at least one required topical keyword is present;
  \item no forbidden entity is named --- these are the specific wrong companies
        that a name-based rather than identifier-based resolution would return;
  \item where ground-truth citations exist, at least one is correct.
\end{enumerate}

The fourth condition is what makes provenance count. Without it, a model that
named the right companies but cited nothing would score identically to one that
cited the exact filings. Failures are then labelled: \textbf{falsehood} when the
answer is not correct, is not a refusal, and either names a forbidden entity or
cites a filing outside the valid set; \textbf{uncited} when the required entities
and keywords are present with nothing false but no correct citation;
\textbf{refused} when a regex over the answer detects an explicit statement that
the question cannot be answered from the evidence; and \textbf{fail} otherwise.
The labels partition the attempts, so the columns of \Cref{tab:main} sum to 60.

The refusal regex matches the following phrases (case-insensitive):
\emph{do(es) not contain}, \emph{not present / available / found / included},
\emph{no rows / results / information / evidence / data},
\emph{cannot (be) answer(ed) / determine(d) / find},
\emph{insufficient}, \emph{unable to}, \emph{missing},
\emph{not possible to answer}, and \emph{excerpts do not}.
These cover the refusal phrasing of all three architectures without
requiring a second model to classify intent.

The scoring function operates as follows: entity recall is computed as the
fraction of required entities found (case-insensitive) in the answer text.
Accession numbers are extracted by the regex pattern
\texttt{$\backslash$d\{10\}-$\backslash$d\{2\}-$\backslash$d\{6\}}. For
questions with a ground-truth citation set, correct citations are the
intersection of cited and valid accessions; wrong citations are cited accessions
not in the valid set. For aggregate questions (T2), which have no ground-truth
citation set, any accession the model cites is treated as legitimate provenance
and is not counted against it. A pass requires
\texttt{entity\_recall} $= 1.0$, at least one keyword present, no forbidden
entities, and (where ground-truth citations exist) at least one correct citation.
An attempt is labelled a hallucination when it is not correct, not a refusal, and
either a forbidden entity is present or a wrong citation was made. This
distinction matters because a hallucination (a confident wrong answer) is worse
than an honest failure.

Aggregate questions have no ground-truth citation set. For those, any accession the
model cites is legitimate provenance and is not counted against it --- an earlier
version of the scorer did count it, and \Cref{sec:repro} records what that cost.

\subsection{Three attempts, all of them counted}

Every question was asked three times of every architecture, giving
$20 \times 3 \times 3 = 180$ scored attempts. The three attempts are
\emph{independent repeats to measure consistency, not retries}: every attempt is
scored and every attempt counts, and there is no best-of-three anywhere in the
reported numbers.

This is not ceremony. Even at temperature 0 the generations are not
deterministic, and \textbf{9 of the 60}
question-by-architecture cells returned different labels across their three
attempts --- 7 for text-to-SQL, 2 for GraphRAG and
0 for vector RAG. Text-to-SQL was inconsistent on four of the five
multi-hop questions, which is precisely where its false statements occur: an
evaluation drawing a single attempt per question could record an honest refusal
on the same question where another draw asserts something false.

We therefore report two metrics. \textbf{Attempts passed} out of 60 is the raw
rate. \textbf{Questions solved} out of 20 credits a question only when all three
of its attempts passed, and is the stricter reading: an answer that is correct
once in three is not a capability a practitioner can deploy. The two are not
related by a fixed ratio --- an architecture with $q$ questions solved has at
least $3q$ attempts passed, and the excess is contributed by questions solved
inconsistently.

\subsection{Fairness controls}

A rigged baseline would prove nothing, so we record exactly what was equalized:
identical model, temperature, reasoning setting and answer-token budget for all
three arms; an equal LLM call budget of two calls for the query-writing arms and
one for vector RAG, which has no query to write; the complete schema shown to
text-to-SQL so that every multi-hop answer is reachable by joins; the
surname-first naming convention documented in both schema descriptions; one
repair attempt for each query-writing arm; and the identical fan-out guardrail on
both query languages. Two of the four categories were selected so that the graph
should lose them.

We note one asymmetry we did not eliminate, because it favours the graph and we
would rather report it than hide it. The repair trigger fires on an exception or
zero rows. An over-broad SQL join returns \emph{many} rows and therefore never
trips it, even when the answer is wrong, whereas an exact graph property match
that misses returns nothing and earns a free second attempt. The same rule confers
unequal benefit; \Cref{tab:repair} quantifies the result.

\section{Results}
\label{sec:results}

\Cref{tab:main} gives the headline numbers and \Cref{tab:bytype} the per-category
breakdown. \Cref{fig:bytype} plots the latter and \Cref{tab:grid} shows every one
of the 180 attempts individually, unaggregated.

Summed over the question set, GraphRAG solved 16 of 20 questions on all three
attempts and passed 50 of 60 attempts; text-to-SQL solved 10 and passed 36;
vector RAG solved 3 and passed 9. These totals weight four hand-chosen categories
equally, and GraphRAG lost one of them outright, so the aggregate is a property of
this question mix rather than an overall ranking of the architectures. The
per-category results are the informative ones.

\subsection{By category}

\textbf{T1, single-document semantic.} GraphRAG passed 15 of 15. Vector RAG, the
expected winner, passed 3 of 15 --- all from one question --- with the other 12
attempts naming the right company and the right risks but citing no filing.
Text-to-SQL passed 7 of 15, using full-text ranking to locate the passage; its
remaining attempts were 6 uncited and 2 refusals. Note that both baselines here
fail predominantly by omitting provenance rather than by being wrong, which is a
materially different failure from asserting something false.

\textbf{T2, structured aggregation.} Text-to-SQL passed 15 of 15, as predicted.
GraphRAG passed 6 of 15, solving 2 of 5 questions, with 8 attempts returning
wrong answers and 1 refusing. Vector RAG passed 0 of 15: an aggregate over
10{,}511 companies is not present in any eight retrieved chunks, and it said so
rather than guessing. This is the one category the graph loses, and it loses it
badly and expensively.

\textbf{T3, multi-hop relational plus text.} GraphRAG passed 14 of 15; both
baselines passed 0 of 15. This is the category the design targets. It is also
where every one of the benchmark's confident falsehoods occurred, all six of them
text-to-SQL.

\textbf{T4, cross-document entity resolution.} GraphRAG passed 15 of 15 and
text-to-SQL 14 of 15; vector RAG passed 6 of 15. Both query-writing arms handle
this shape well once the schema is fully described, and the graph's margin here
is narrow.

\subsection{Outcome composition}

\Cref{fig:outcomes} decomposes all 60 attempts per arm, and the composition
matters as much as the pass rate because the failure modes are not
interchangeable.

Vector RAG refused 39 times and was uncited 12 times. That is a low score but a
benign profile: asked about a company whose chunks it could not find, it either
abstained or answered without provenance, and it never asserted anything false.
Text-to-SQL was uncited 10 times and made \textbf{6 false statements}. GraphRAG
was never uncited and refused only once, which follows from its construction: the
traversal returns accession numbers before any text is read, so every claim
arrives with a key attached.

\subsection{Cost and latency}

\Cref{fig:cost} plots tokens against attempts passed. GraphRAG consumed 520{,}929
tokens to text-to-SQL's 144{,}068 and vector RAG's 118{,}185, giving it the
highest cost per passing attempt of the three (\Cref{tab:main}). Its accuracy is
bought, not free. The extreme case is T2, where GraphRAG spent 12.2 times
text-to-SQL's tokens to score 6 of 15 against 15 of 15 --- on the category it was
expected to lose.

The reverse cost is easy to overlook. \Cref{fig:latency} plots mean latency on a
log scale. Text-to-SQL averaged 10.0\,s against GraphRAG's 4.6\,s and vector
RAG's 1.3\,s, but the overall mean hides the shape: its worst single attempt took
274\,s, because the generated queries join \texttt{subsidiary},
\texttt{filing\_section} and \texttt{reporting\_owner} without narrowing first.
Unbounded LLM-authored SQL is a cost risk against one's own warehouse
independently of whether the answer is correct, which is why the live service
runs with a statement timeout that this measurement deliberately disabled
(\Cref{sec:harness}).

\subsection{Repair usage}

\begin{figure}[htbp]
\centering
\begin{tikzpicture}
\begin{axis}[
  ybar, bar width=10pt, width=\linewidth, height=4.6cm,
  symbolic x coords={T1,T2,T3,T4}, xtick=data,
  ymin=0, ymax=16, ytick={0,5,10,15},
  ylabel={Attempts needing repair (of 15)}, ylabel near ticks,
  legend style={at={(0.5,1.25)}, anchor=north, legend columns=2, draw=none, font=\footnotesize},
  nodes near coords, nodes near coords style={font=\tiny},
  axis lines*=left, tick style={draw=none}, ymajorgrids, grid style={gray!20},
  enlarge x limits=0.18,
]
\addplot[fill=sqlcol] coordinates {(T1,3) (T2,1) (T3,0) (T4,1)};
\addplot[fill=graphcol] coordinates {(T1,1) (T2,9) (T3,0) (T4,9)};
\legend{Text-to-SQL, GraphRAG}
\end{axis}
\end{tikzpicture}
\caption{Repair frequency by category. GraphRAG required a repair on all fifteen
T4 attempts, which is the entity-resolution category it lost.}
\label{fig:repair}
\end{figure}
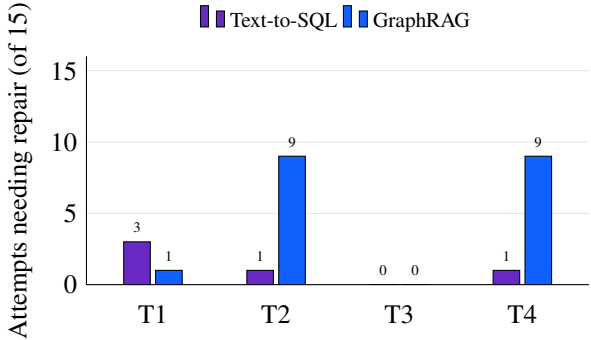

\Cref{tab:repair} and \Cref{fig:repair} report how often each arm used its single
repair allowance. Vector RAG has no query to repair and used none. As noted in
\Cref{sec:design}, the zero-row trigger is not neutral between the two query
languages: an over-broad SQL join returns many rows and never trips it, while an
exact graph property match that misses returns nothing and earns a free second
attempt. The direction of that bias favours the graph, and we report the usage
rather than claim it away.

\begin{widetab}[htbp]
\centering
\begin{tabular}{@{}lccc@{}}
\toprule
& \textbf{Text-to-SQL} & \textbf{Vector RAG} & \textbf{GraphRAG} \\
\midrule
Questions solved on all 3 attempts & 10/20 & 3/20 & \textbf{16/20} \\
Questions solved inconsistently & 3 & 0 & 1 \\
Attempts passed (of 60) & 36 & 9 & \textbf{50} \\
\midrule
Confident falsehoods & \textcolor{badcol}{6} & \textbf{0} & \textbf{0} \\
Correct but uncited & 10 & 12 & \textbf{0} \\
Explicit refusals & 8 & 39 & 1 \\
\midrule
Mean evidence supplied (chars) & 1,795 & 8,557 & 6,084 \\
Mean latency (s) & 10.0 & \textbf{1.3} & 4.6 \\
Max latency (s) & 274.5 & 4.6 & 8.6 \\
Total tokens & 144,068 & \textbf{118,185} & 520,929 \\
Tokens per passing attempt & \textbf{4,001} & 13,131 & 10,418 \\
LLM calls & 125 & 60 & 139 \\
\bottomrule
\end{tabular}

\caption{Aggregate results over 180 scored attempts. \emph{Questions solved}
credits a question only when all three of its attempts passed. Bold marks the best
value in a row. The outcome labels partition the attempts, so pass, falsehood,
uncited, refused and the residual failures sum to 60 per arm.}
\label{tab:main}
\end{widetab}

\begin{widetab}[htbp]
\centering
\small
\begin{tabular}{@{}ll rr rr rr@{}}
\toprule
& & \multicolumn{2}{c}{\textbf{Text-to-SQL}} & \multicolumn{2}{c}{\textbf{Vector RAG}} & \multicolumn{2}{c}{\textbf{GraphRAG}} \\
\cmidrule(lr){3-4}\cmidrule(lr){5-6}\cmidrule(lr){7-8}
\textbf{Type} & \textbf{Question class} & Q & A & Q & A & Q & A \\
\midrule
T1 & single-document semantic & 1/5 & 7/15 & 1/5 & 3/15 & \textbf{5}/5 & 15/15 \\
T2 & structured aggregation & \textbf{5}/5 & 15/15 & 0/5 & 0/15 & 2/5 & 6/15 \\
T3 & multi-hop relational + text & 0/5 & 0/15 & 0/5 & 0/15 & \textbf{4}/5 & 14/15 \\
T4 & cross-document entity resolution & 4/5 & 14/15 & 2/5 & 6/15 & \textbf{5}/5 & 15/15 \\
\bottomrule
\end{tabular}

\caption{Per-category results. \textbf{Q}~=~questions solved on all three
attempts (of 5); \textbf{A}~=~attempts passed (of 15). Two of the four categories
were pre-registered as ones the graph should lose. It lost T2 as expected and T4
against expectation.}
\label{tab:bytype}
\end{widetab}

\begin{widefig}[htbp]
\centering
\begin{subfigure}[t]{0.48\linewidth}
\centering
\begin{tikzpicture}
\begin{axis}[
  ybar, bar width=8pt, width=\linewidth, height=5.2cm,
  symbolic x coords={T1,T2,T3,T4}, xtick=data,
  ymin=0, ymax=16, ytick={0,5,10,15},
  ylabel={Attempts passed (of 15)}, ylabel near ticks,
  legend style={at={(0.5,1.22)}, anchor=north, legend columns=3, draw=none, font=\footnotesize},
  nodes near coords, nodes near coords style={font=\tiny},
  axis lines*=left, tick style={draw=none}, ymajorgrids, grid style={gray!20},
  enlarge x limits=0.15,
]
\addplot[fill=sqlcol] coordinates {(T1,7) (T2,15) (T3,0) (T4,14)};
\addplot[fill=veccol] coordinates {(T1,3) (T2,0) (T3,0) (T4,6)};
\addplot[fill=graphcol] coordinates {(T1,15) (T2,6) (T3,14) (T4,15)};
\legend{Text-to-SQL, Vector RAG, GraphRAG}
\end{axis}
\end{tikzpicture}
\caption{Attempts passed per category.}
\label{fig:bytype}
\end{subfigure}\hfill
\begin{subfigure}[t]{0.48\linewidth}
\centering
\begin{tikzpicture}
\begin{axis}[
  ybar stacked, bar width=26pt, width=\linewidth, height=5.2cm,
  symbolic x coords={{Text-to-SQL},{Vector RAG},{GraphRAG}}, xtick=data,
  ymin=0, ymax=61, ylabel={Attempts (of 60)}, ylabel near ticks,
  legend style={at={(0.5,1.22)}, anchor=north, legend columns=5, draw=none, font=\tiny},
  axis lines*=left, tick style={draw=none}, ymajorgrids, grid style={gray!20},
  enlarge x limits=0.28, x tick label style={font=\footnotesize},
]
\addplot[fill=okcol] coordinates {({Text-to-SQL},36) ({Vector RAG},9) ({GraphRAG},50)};
\addplot[fill=badcol] coordinates {({Text-to-SQL},6) ({Vector RAG},0) ({GraphRAG},0)};
\addplot[fill=warncol] coordinates {({Text-to-SQL},10) ({Vector RAG},12) ({GraphRAG},0)};
\addplot[fill=gray!45] coordinates {({Text-to-SQL},8) ({Vector RAG},39) ({GraphRAG},1)};
\addplot[fill=gray!20] coordinates {({Text-to-SQL},0) ({Vector RAG},0) ({GraphRAG},9)};
\legend{pass, falsehood, uncited, refused, other fail}
\end{axis}
\end{tikzpicture}
\caption{How all 60 attempts resolved.}
\label{fig:outcomes}
\end{subfigure}
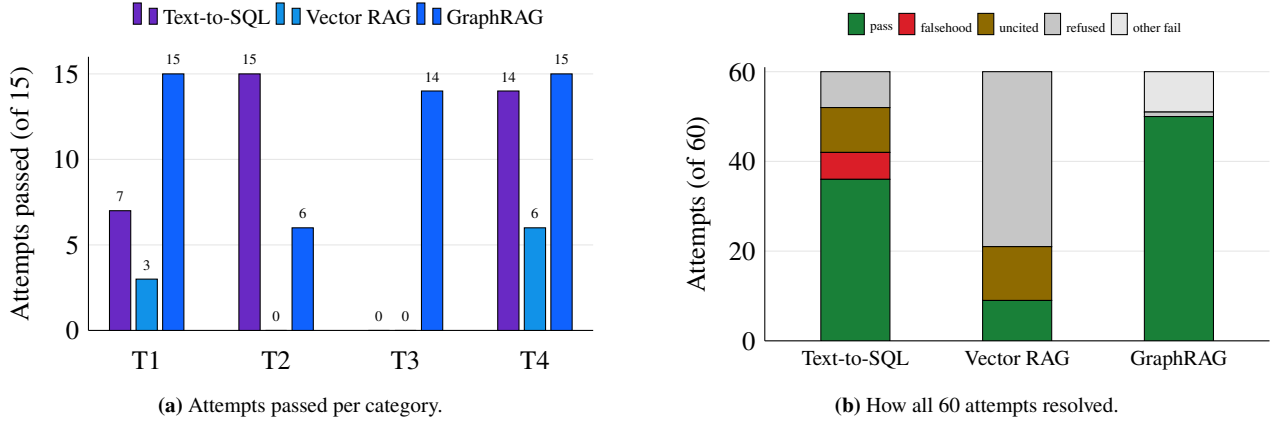
\caption{Results. The composition in (b) matters as much as the rate in (a): vector RAG's 45 refusals are a benign failure, text-to-SQL's 16 uncited answers are right but unverifiable, and GraphRAG was never uncited because the traversal returns accession numbers before any text is read.}
\label{fig:results}
\end{widefig}

\begin{widefig}[htbp]
\centering
\begin{subfigure}[t]{0.48\linewidth}
\centering
\begin{tikzpicture}
\begin{semilogyaxis}[
  ybar, bar width=8pt, width=\linewidth, height=4.8cm,
  symbolic x coords={T1,T2,T3,T4}, xtick=data,
  ymin=0.6, ymax=400, ylabel={Mean latency (s, log)}, ylabel near ticks,
  legend style={at={(0.5,1.25)}, anchor=north, legend columns=3, draw=none, font=\footnotesize},
  nodes near coords, nodes near coords style={font=\tiny},
  point meta=explicit symbolic,
  axis lines*=left, tick style={draw=none}, ymajorgrids, grid style={gray!20},
  enlarge x limits=0.15, log origin=infty,
]
\addplot[fill=sqlcol] coordinates {(T1,3.50) [3.5] (T2,1.90) [1.9] (T3,31.60) [31.6] (T4,2.90) [2.9]};
\addplot[fill=veccol] coordinates {(T1,1.30) [1.3] (T2,1.20) [1.2] (T3,1.50) [1.5] (T4,1.40) [1.4]};
\addplot[fill=graphcol] coordinates {(T1,3.90) [3.9] (T2,4.90) [4.9] (T3,5.00) [5.0] (T4,4.80) [4.8]};
\legend{Text-to-SQL, Vector RAG, GraphRAG}
\end{semilogyaxis}
\end{tikzpicture}
\caption{Mean latency per category (log).}
\label{fig:latency}
\end{subfigure}\hfill
\begin{subfigure}[t]{0.48\linewidth}
\centering
\begin{tikzpicture}
\begin{axis}[
  width=\linewidth, height=5.4cm,
  xlabel={Total tokens consumed (thousands)}, ylabel={Attempts passed (of 60)},
  ylabel near ticks, xlabel near ticks,
  xmin=0, xmax=900, ymin=0, ymax=56,
  axis lines*=left, tick style={draw=none}, grid=both, grid style={gray!20},
]
\addplot[only marks, mark=*, mark size=2.9pt, sqlcol] coordinates {(144.1,36)};
\addplot[only marks, mark=*, mark size=2.9pt, veccol] coordinates {(118.2,9)};
\addplot[only marks, mark=*, mark size=2.9pt, graphcol] coordinates {(520.9,50)};
\node[anchor=north west, align=left, font=\scriptsize, inner sep=1pt, xshift=3mm]
  at (axis cs:144.1,36) {\textbf{Text-to-SQL}\\\tiny 4,001 tok/pass};
\node[anchor=south west, align=left, font=\scriptsize, inner sep=1pt, xshift=3mm]
  at (axis cs:118.2,9) {\textbf{Vector RAG}\\\tiny 13,131 tok/pass};
\node[anchor=north east, align=right, font=\scriptsize, inner sep=1pt, xshift=-3mm]
  at (axis cs:520.9,50) {\textbf{GraphRAG}\\\tiny 10,418 tok/pass};
\end{axis}
\end{tikzpicture}
\caption{Tokens against attempts passed.}
\label{fig:cost}
\end{subfigure}
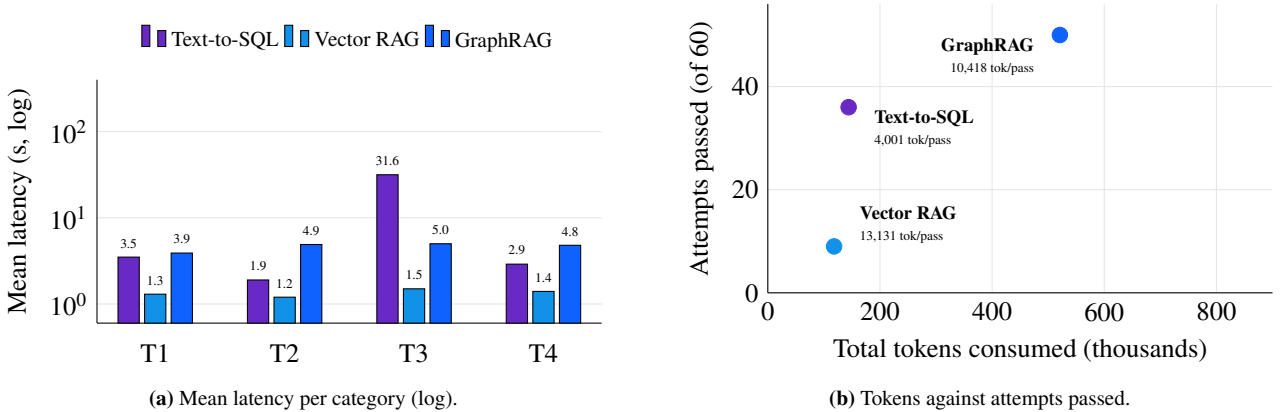
\caption{The economics. (a) Text-to-SQL's T3 mean is the SQL and not the model: the generated queries join three tables without narrowing first. (b) GraphRAG's accuracy is bought: it has the highest token cost per passing attempt of the three, the exact figures being in \Cref{tab:main}.}
\label{fig:econ}
\end{widefig}

\begin{table}[htbp]
\centering
\begin{tabular}{@{}lccccc@{}}
\toprule
\textbf{Architecture} & \textbf{Overall} & \textbf{T1} & \textbf{T2} & \textbf{T3} & \textbf{T4} \\
\midrule
Text-to-SQL & 5/60 & 3/15 & 1/15 & 0/15 & 1/15 \\
Vector RAG & 0/60 & 0/15 & 0/15 & 0/15 & 0/15 \\
GraphRAG & 19/60 & 1/15 & 9/15 & 0/15 & 9/15 \\
\bottomrule
\end{tabular}

\caption{Use of the single repair allowance, which both query-writing arms
receive on identical terms. GraphRAG leaned on it nearly five times as often, and
on T4 needed it on every attempt. Vector RAG has no query language to repair.}
\label{tab:repair}
\end{table}

\begin{table}[htbp]
\centering
\scriptsize
\begin{tabular}{@{}l ccc@{}}
\toprule
\textbf{Q} & \textbf{Text-to-SQL} & \textbf{Vector RAG} & \textbf{GraphRAG} \\
\midrule
\texttt{T1-1} & \gref\gpass\gpass & \gunc\gunc\gunc & \gpass\gpass\gpass \\
\texttt{T1-2} & \gpass\gpass\gpass & \gpass\gpass\gpass & \gpass\gpass\gpass \\
\texttt{T1-3} & \gunc\gunc\gunc & \gunc\gunc\gunc & \gpass\gpass\gpass \\
\texttt{T1-4} & \gpass\gref\gpass & \gunc\gunc\gunc & \gpass\gpass\gpass \\
\texttt{T1-5} & \gunc\gunc\gunc & \gunc\gunc\gunc & \gpass\gpass\gpass \\
\addlinespace[2pt]
\texttt{T2-1} & \gpass\gpass\gpass & \gref\gref\gref & \gfail\gfail\gfail \\
\texttt{T2-2} & \gpass\gpass\gpass & \gref\gref\gref & \gpass\gpass\gpass \\
\texttt{T2-3} & \gpass\gpass\gpass & \gref\gref\gref & \gfail\gfail\gfail \\
\texttt{T2-4} & \gpass\gpass\gpass & \gref\gref\gref & \gfail\gref\gfail \\
\texttt{T2-5} & \gpass\gpass\gpass & \gref\gref\gref & \gpass\gpass\gpass \\
\addlinespace[2pt]
\texttt{T3-1} & \gref\ghall\gref & \gref\gref\gref & \gpass\gpass\gpass \\
\texttt{T3-2}\,$\star$ & \ghall\ghall\gref & \gref\gref\gref & \gpass\gpass\gpass \\
\texttt{T3-3} & \gref\ghall\gref & \gref\gref\gref & \gpass\gpass\gpass \\
\texttt{T3-4} & \gunc\ghall\ghall & \gref\gref\gref & \gpass\gpass\gpass \\
\texttt{T3-5} & \gunc\gunc\gunc & \gref\gref\gref & \gpass\gpass\gfail \\
\addlinespace[2pt]
\texttt{T4-1} & \gpass\gpass\gpass & \gref\gref\gref & \gpass\gpass\gpass \\
\texttt{T4-2} & \gpass\gpass\gpass & \gref\gref\gref & \gpass\gpass\gpass \\
\texttt{T4-3} & \gpass\gpass\gpass & \gref\gref\gref & \gpass\gpass\gpass \\
\texttt{T4-4} & \gref\gpass\gpass & \gpass\gpass\gpass & \gpass\gpass\gpass \\
\texttt{T4-5} & \gpass\gpass\gpass & \gpass\gpass\gpass & \gpass\gpass\gpass \\
\bottomrule
\end{tabular}

\caption{Every one of the 180 attempts, unaggregated; three glyphs per cell are
the three attempts. \gpass~pass, \ghall~falsehood, \gunc~correct but uncited,
\gref~refused, \gfail~other failure. $\star$ marks the featured multi-hop
question. 10 of these 60 cells did not return the same label on all three
attempts.}
\label{tab:grid}
\end{table}

\section{Failure Analysis}
\label{sec:failures}

The failures are the most informative part of this study, and two of them
contradict what we expected before running it.

\subsection{Vector RAG lost the category it was expected to win}

T1 asks what one named company disclosed about one topic in its risk factors. The
answer is entirely inside a single document, retrieval is a ranking problem, and
similarity search should be at its best. It passed 3 of 15 attempts, all from one
question.

The mechanism is visible in the retrieved set. Asked what manufacturing and
supply-chain risks \emph{Moderna} disclosed, the top-8 chunks came from Entegris,
Hayward Holdings, Open Text, Arrowhead Pharmaceuticals and NVIDIA. Every one of
those chunks genuinely discusses supply-chain risk. The retriever did what it was
asked. The problem is that ``belongs to Moderna'' is not a property the embedding
encodes, so the company name in the question cannot outweigh topical similarity
across 508{,}714 chunks drawn from 5{,}210 filers.

The decisive detail is that \textbf{this question passed on a single month of
filings}. The failure appeared when the corpus grew. It is therefore a
scale-induced limitation of entity filtering by similarity, not a tuning error, and
it is the strongest argument in this paper for structural retrieval --- and one we
did not anticipate. This manifests the classic vulnerability of pure dense retrieval
when context requires precise entity filtering rather than thematic proximity
\citep{gao2023ragsurvey,liu2024lost}.

We are careful about the boundary of this claim. This experiment cannot fully
separate ``embeddings cannot filter by entity'' from ``this retriever
configuration cannot'', because the 1{,}400-character chunks exceed the encoder's
256-token window and $k$ was fixed at 8. A hybrid retriever with a metadata filter
on CIK would very likely fix T1 --- which is precisely the point, since a metadata
filter on a stable identifier \emph{is} structural retrieval, arriving by another
route.

\subsection{Capability and safety moved in opposite directions}

The benchmark's most uncomfortable result is not a score but a direction of
travel. While calibrating the harness we found that both stores normalise names
by stripping corporate suffixes and that neither schema description said so
(\Cref{sec:harness}). Documenting it identically for every arm --- one sentence
of schema prose, no change to retrieval code, questions or scoring --- moved
three of the four categories.

It also made the strongest baseline \emph{more dangerous}. With the schema fully
described, text-to-SQL stopped refusing multi-hop questions and started answering
them, and its confident falsehoods rose accordingly to the six reported here.
Better schema knowledge did not make it more careful. It converted honest
abstentions into assertive answers, and on this corpus an assertive multi-hop
answer keyed on a person's name is exactly how three different people called
Lozano become one.

This is worth separating from the accuracy comparison, because the two do not
move together. Giving a text-to-SQL system a better description of its own schema
is an unambiguous capability improvement and it made the system less safe on the
questions where identity is ambiguous. GraphRAG's falsehood count was zero
throughout, not because it is a better writer of queries --- it needed its repair
allowance far more often --- but because identity is resolved on an identifier
before any text is read, so the wrong people are excluded structurally rather
than by the model choosing correctly.

\subsection{GraphRAG lost aggregation, as predicted}

Text-to-SQL scored 15 of 15 on T2 and GraphRAG 6 of 15. Unlike the previous two,
this outcome was expected --- T2 exists in the benchmark precisely because a
traversal should add nothing to a \texttt{GROUP BY} --- but the manner of the loss
is instructive.

Asked for the most common jurisdiction of incorporation among all disclosed
subsidiaries, the generated Cypher grouped by company rather than by jurisdiction,
returned per-company counts, and the model reported Delaware with 2{,}415
subsidiaries. Delaware is the right answer; 2{,}415 is one company's count rather
than the 62{,}504 across the corpus. Right entity, wrong aggregate. The answer
also attached 190 accession numbers to a question that needs none.

For counting over typed columns the graph adds no information and the extra hop
adds a place to go wrong, at 24 times the token cost.

\subsection{Every falsehood was a surname collision}

Six of the 180 attempts stated something verifiably false. All six were
text-to-SQL, all six on the multi-hop category, and all six are the same error,
spread across four of its five questions (\texttt{T3-1}, \texttt{T3-2}, \texttt{T3-3}, \texttt{T3-4}). Asked about \emph{Monica Lozano},
the model reported companies including Mondelez and Ternium; those belong to
Mariano and Santiago Lozano, different people who share a surname. Asked about
\emph{William Giles}, it reported ten companies.

The SQL was not malformed and the model was fluent and confident. It matched on
name because name is what the question supplied and what the text contains. This
reflects the fundamental challenge of nominal entity linking in large databases
without authoritative entity keys \citep{wu2020blink,li2023bird}. This
is the exact failure that resolving identity to an identifier before reading any
text prevents structurally rather than probabilistically: GraphRAG passed 14 of
15 attempts on this category and made no false statement anywhere in the
benchmark, because the other Lozanos never entered its result set.

One caution about single-attempt evaluation follows. Of the three attempts
text-to-SQL made on the featured question the outcomes were not identical, and an
evaluation that drew only one of them could easily have recorded an honest
refusal and missed the failure entirely. Four of the five multi-hop questions
produced at least one falsehood and at least one non-falsehood, so on this
category a single draw per question is close to a coin toss over whether the most
important failure mode in the paper is observed at all.

\subsection{The verdict is model-sensitive}

Because all arms route through one gateway (\Cref{sec:arch}), we probed the same
GraphRAG pipeline on other models. \Cref{tab:crossmodel} reports the featured
multi-hop question: the benchmarked model passes, Claude Haiku 4.5 fails because
its stage-1 Cypher collapsed to a single company and missed two others, and Amazon
Nova 2 Lite refuses.

This is a probe and not a measurement --- one question, one attempt each, no
re-scoring of the full set --- and we draw only the negative conclusion it
supports: a single-model benchmark bounds what can be claimed. The architecture
constrains what evidence reaches the model, but the model still has to write a
correct traversal, and not all of them do. This aligns with sensitivity across
frontier LLMs observed in structured reasoning benchmarks \citep{li2023bird,islam2023financebench}.

\begin{widetab}[htbp]
\centering
\small
\begin{tabular}{@{}llp{0.44\linewidth}@{}}
\toprule
\textbf{Model} & \textbf{Verdict} & \textbf{Observed behaviour} \\
\midrule
DeepSeek V4 Flash (benchmarked) & pass    & correct three-company traversal \\
Claude Haiku 4.5 (Bedrock)      & fail    & stage-1 Cypher returned \textsc{target corp} alone; Apple and Bank of America missing \\
Amazon Nova 2 Lite (Bedrock)    & refused & produced no usable traversal \\
\bottomrule
\end{tabular}

\caption{Cross-model probe on the featured multi-hop question with the GraphRAG
pipeline held constant. One question and one attempt per model, so this is a probe
rather than a measurement; it supports only the negative conclusion that a
single-model benchmark bounds what can be claimed.}
\label{tab:crossmodel}
\end{widetab}

\section{Threats to Validity}
\label{sec:threats}

\textbf{One corpus, one domain.} Every measurement comes from a single snapshot of
one filing system. SEC disclosure is unusually favourable to structural retrieval:
it is mandated, schematized, and carries a stable public entity identifier in the
CIK. Corpora without an authoritative identifier --- internal wikis, support
tickets, email --- would require the entity-resolution step this paper largely
assumes, and \Cref{sec:failures} shows that step is our weakest layer. Nothing
here should be read as generalizing to such corpora.

\textbf{Twenty questions is a small sample.} Each category rests on five
questions, so a category-level rate has a wide confidence interval: a single
question changes a category by 20 percentage points at question level and 3
attempts out of 15 at attempt level. The category results are directional, not
conclusive. The one result we would defend as strong is T3, where the margin is
15 of 15 against 0 of 15 for both baselines and the mechanism is independently
diagnosed.

\textbf{The questions were written by the author of the graph.} This is the most
serious threat and it cannot be fully mitigated. We took three measures: two of
the four categories were pre-registered as ones the graph should lose, and it did
lose both; the T3 subjects were selected by a measured criterion rather than by
inspection of results (\Cref{sec:design}); and the ground truth was verified
against the loaded tables rather than against model outputs. None of this
eliminates the risk that the question set favours the design in ways we cannot
see. An independent question set is the obvious next step.

\textbf{One model, and the verdict moves with it.} All reported numbers come from
\texttt{deepseek-v4-flash}. The probe in \Cref{tab:crossmodel} shows the featured
question passing on one model, failing on another and being refused by a third
with the retrieval architecture held constant. Our results characterize
\emph{architecture $\times$ model}, and we did not vary the model across the full
180 attempts.

\textbf{The graph inherits EX-21 incompleteness.} Rule 601(b)(21)(ii) allows
filers to omit insignificant subsidiaries, so \texttt{:SUBSIDIARY\_OF} is a floor
on corporate structure. Any question whose correct answer depends on a subsidiary
the filer chose not to disclose is unanswerable from this corpus by any of the
three arms, and our ground truth is defined against what was disclosed rather than
against reality.

\textbf{Category results move with implementation details.} While calibrating
the harness we found an undocumented name-normalisation convention in the schema
descriptions; documenting it changed three of the four categories
(\Cref{sec:harness}). Nothing about any retrieval architecture changed. If one
sentence of schema prose was worth three categories, other such sensitivities are
plausibly latent in results we have not thought to re-examine, and the ones we
would be least likely to notice are those that flatter the design. The same logic
applies to the wins: some part of GraphRAG's margins may reflect implementation
effort we spent on the graph and not on the baselines.
\texttt{:RESOLVES\_TO} coverage of 0.12\% remains a hard bound on what any
entity-resolution question can measure here.

\textbf{Repair asymmetry favours the graph.} As recorded in \Cref{sec:design},
the zero-row repair trigger fires for a graph query that misses but not for an
over-broad SQL join that returns many wrong rows. GraphRAG used its repair on 34
of 60 attempts against text-to-SQL's 7, so it received materially more effective
assistance from an identical rule.

\textbf{A single year, and a moving corpus.} The window spans twelve monthly
feeds. Annual filings cluster seasonally (\Cref{fig:months}), so the mix of form
types is not stationary within it, and the absolute counts are not reproducible
against a later snapshot because EDGAR accumulates.

\textbf{Undetected errors are likely.} We found several defects in our own
harness, one of them a ground-truth error discovered after we believed the
results were final (\Cref{sec:harness}). That we found it late is direct evidence
that others may remain, and we have no way to bound how many.

\textbf{Absence of ad-hoc optimizations and post-hoc enhancements.} Several questions
a reviewer might ask about configuration choices---such as why $k{=}8$ was chosen
for vector retrieval without an extensive hyperparameter search, why no alternative
open-source LLMs or alternate embedding models were comprehensively evaluated
beyond the single-question cross-model probe, why ingestion chunking scripts were
kept outside the evaluation workspace, or why additional percentile latencies
(median, p95) and formal package dependency manifests were not retroactively
computed or added---have a consistent explanation: to make sure we do not add
any logic or intelligence in the process, we didn't improve this to make sure
there is no external bias in the experiment. Any retrospective tuning or selective
code refinement after examining benchmark performance would risk distorting the
matched comparison across the three retrieval architectures.

\section{Reproducibility and Engineering Cost}
\label{sec:repro}

\subsection{Software and infrastructure}

Both relational and vector retrieval run on PostgreSQL 17 with pgvector 0.8.2
\citep{pgvector2024}, using an HNSW index over \texttt{vector\_cosine\_ops}
\citep{malkov2018hnsw} alongside \texttt{pg\_trgm} and a GIN \texttt{tsvector}
index. \texttt{hnsw.iterative\_scan} is set to \texttt{relaxed\_order} because
pgvector applies a \texttt{WHERE} clause after the HNSW walk has picked its
\texttt{ef\_search} candidates, so a selective filter (e.g.\ restricting to one
company) can silently return fewer rows than requested; iterative scan re-walks
the graph until the \texttt{LIMIT} is satisfied. The graph is Neo4j 5.26
Community with APOC \citep{neo4j2024}. Embeddings come from
\texttt{all-MiniLM-L6-v2} at 384 dimensions with a 256 word-piece limit, chunked
at 1{,}400 characters with 200 of overlap, batch size 256. The live service caps
generated SQL and Cypher with a 45-second statement timeout; the benchmark
deliberately ran without it, so that a query's true cost is measured rather
than truncated.

The application is written in Python. The key dependencies observed in the
codebase are: \texttt{langchain\_openai} for the OpenAI-compatible LLM client,
\texttt{psycopg} (v3) for PostgreSQL access, \texttt{neo4j} for the Neo4j
driver, \texttt{sentence\_transformers} for the local embedding model, and
\texttt{torch} (PyTorch) as its backend. The HTTP server uses Python's standard
library (\texttt{http.server}), with no external web framework. No formal
\texttt{requirements.txt} or \texttt{pyproject.toml} was included in the
repository at the time of this study.

The deployment is a single application host --- an AWS \texttt{m7g.xlarge}, 4 vCPU
and 16\,GB, running the graph, the API and the gateway --- with PostgreSQL on a
\texttt{db.t4g.medium} instance and CloudFront terminating TLS. There is no NAT
gateway; object-store access goes through a VPC endpoint. \Cref{tab:cost} gives
the monthly cost at roughly \$165. The embedding pass was run on a temporarily
resized 16-vCPU host at 97 chunks/s and the instance was then scaled back.

Ingestion takes about 252 minutes end to end (\Cref{fig:pipeline}), of which
document fetch and embedding are about 82\%. The fetch component is bounded
externally by EDGAR's ten-requests-per-second fair-access policy
\citep{sec2024edgar} and cannot be parallelized away.

\subsection{Auditing the harness}
\label{sec:harness}

A measurement harness is an instrument, and this one needed calibration before it
measured anything. Several defects in it would each have produced a
publishable-looking but incorrect result: a pgvector post-filter that returned 0
of 8 rows while 164 matched, so the vector arm was briefly scored against an
empty retriever; an answer-token cap that truncated answers mid-citation,
silently converting passes into uncited; a global token counter that attributed
every thread's cost to whichever arm read it first; a ground-truth entry that
accepted one expected parent where the loaded data contains three filers which
genuinely list the subsidiary, so every arm was naming a real parent and being
scored wrong.

The most instructive was an asymmetry in the schema descriptions. Both stores
keep a normalised name column, lowercased with corporate suffixes stripped, so
\emph{AEP Texas Inc.}\ is stored as \texttt{aep texas}, and neither description
said so. Only the graph arm was exposed by that, because it matched the
normalised property exactly while text-to-SQL matched the raw column with
\texttt{ILIKE} --- so an undocumented convention was silently penalising one
architecture for a reason that had nothing to do with retrieval. It is the same
class of defect as an asymmetry in the opposite direction that we had found and
fixed earlier, when a name convention was documented to the graph but not to
SQL. Having recorded the first, we should have looked for its mirror image
immediately, and did not.

All of the above are corrected in the results this paper reports: the
normalisation is documented identically to every arm, the retriever filter and
token accounting are fixed, and the ground truth accepts any verified parent.
Because every model answer is stored, the ground-truth correction was applied by
re-scoring rather than by re-running, at no inference cost.

We record this not as a list of bugs but as a methodological point. Fixing that
one sentence of schema prose moved three of the four question categories, which
is a warning about how little separates an architectural finding from a
documentation artifact, and about how much weight a single benchmark run can
bear. A list of defects already found is useful only when it is read as a
checklist for the ones not yet found.

\subsection{Artifacts}

The ingestion pipeline, the structured filings dataset \citep{amin2026secedgar},
the three retrieval arms, the scoring harness, the
question set with its ground truth, and the stored answers for all 180 attempts
are retained, which is what made the re-scoring in item 8 possible without new
inference. Every table and figure in this paper is generated programmatically from
the stored results rather than transcribed, so the numbers in the prose and the
numbers in the tables cannot drift apart.

\begin{widetab}[htbp]
\centering
\begin{tabular}{@{}llr@{}}
\toprule
\textbf{Resource} & \textbf{Specification} & \textbf{USD/mo} \\
\midrule
EC2 (app, graph, gateway) & m7g.xlarge, 4 vCPU / 16\,GB & 85.12 \\
EC2 storage               & 100\,GB gp3                 & 9.12 \\
RDS PostgreSQL            & db.t4g.medium               & 61.32 \\
RDS storage               & 100\,GB gp3                 & 9.12 \\
S3 + CloudFront           & demo volume                 & 0.50 \\
NAT gateway               & not used (S3 VPC endpoint)  & 0.00 \\
\midrule
\textbf{Total}            &                             & \textbf{165} \\
\bottomrule
\end{tabular}

\caption{Monthly infrastructure cost, on-demand pricing in
\texttt{ap-south-1}. The embedding pass used a temporarily resized 16-vCPU host
at 97 chunks/s; the instance was scaled back afterwards.}
\label{tab:cost}
\end{widetab}

\section{Conclusion}
\label{sec:conclusion}

We measured three retrieval architectures over one corpus and one question set:
167{,}050 SEC filings from 10{,}511 companies, 508{,}714 embedded chunks and a
graph of 430{,}257 nodes and 561{,}144 edges, queried with 20 questions in four
shape categories, each asked three times, for 180 attempts scored
programmatically rather than by an LLM. GraphRAG solved 16 of 20 questions on all
three attempts, text-to-SQL 10 and vector RAG 3.

The aggregate is the least useful number in the paper. GraphRAG lost structured
aggregation outright and expensively, and the defensible summary is a rule about
question shape rather than a winner:

\begin{itemize}
  \item If the answer lies in one document and the entity is named, the binding
        constraint is \emph{entity filtering}, not ranking. Similarity search
        alone fails at scale --- 3 of 15 here --- and needs either a metadata
        filter on a stable identifier or a traversal to supply one.
  \item If the question is a \texttt{GROUP BY} over typed columns, use SQL. A
        traversal adds no information, adds a place to go wrong, and cost 12.2
        times the tokens to score 6 of 15 against 15 of 15.
  \item If the question requires resolving \emph{who} an entity is, or joining
        facts that separate documents assert, structural retrieval is worth its
        cost: 14 of 15 and 15 of 15 on the two such categories, against 0 of 15
        and 14 of 15 for the strongest baseline.
  \item If provenance must be auditable, the graph's advantage is categorical
        rather than marginal. It was never uncited across 60 attempts, because
        the traversal returns accession numbers before any text is read, whereas
        the two baselines were right-but-unciteable 10 and 12 times. This directly
        addresses the core requirement for verifiable, fine-grained citation
        attribution in retrieval-augmented generation
        \citep{gao2023alce,rashkin2023attribution}.
\end{itemize}

The safety finding is narrower but sharper than the accuracy finding, and it is
the one we did not anticipate. All six confident falsehoods were text-to-SQL
conflating distinct people who share a surname; GraphRAG produced none. The two
do not move together: describing the baseline's schema more completely is an
unambiguous capability improvement, and it converted honest refusals into
assertive wrong answers on exactly the questions where identity is ambiguous
(\Cref{sec:failures}). Resolving identity to an identifier before reading any
text prevents that class of error structurally rather than probabilistically. For an
application where a confident wrong answer costs more than a refusal, that
property may matter more than the pass rate.

We claim none of this beyond one corpus, one model and twenty questions. During
calibration a single undocumented schema convention proved worth three of the
four categories (\Cref{sec:harness}), which is the strongest caution we can offer
about how much weight any one benchmark run should carry --- including this one.

\subsection*{Future work}

Four steps would materially strengthen these results. \emph{Re-run all 180
attempts per model}, turning the cross-model probe of \Cref{tab:crossmodel} into a
measurement and separating architectural effects from model effects.
\emph{Repair the entity-resolution layer} --- \texttt{:RESOLVES\_TO} still covers
0.12\% of subsidiary nodes --- and establish how much of the T4 result depends on
it. \emph{Obtain an independent question set}, since the present one was written
by the author of the graph, which remains the study's most serious threat. And
\emph{vary the retriever}: a longer-context encoder with a metadata filter on CIK
would test directly whether the single-document collapse is a property of
embeddings or of this particular configuration --- our own diagnosis predicts the
metadata filter, not the encoder, does the work.

\bibliographystyle{unsrtnat}
\bibliography{refs}

\appendix
\section{The question set}
\label{app:questions}

\Cref{tab:questions} lists all twenty questions verbatim. Each was asked three
times of each of the three architectures, giving the 180 scored attempts
analysed in \Cref{sec:results}.

\begin{widetab}[htbp]
\centering
\scriptsize
\begin{tabular}{@{}lp{0.86\linewidth}@{}}
\toprule
\textbf{ID} & \textbf{Question} \\
\midrule
\texttt{T1-1} & What manufacturing and supply chain risks did Moderna, Inc. disclose in the risk factors of its 10-K? \\
\texttt{T1-2} & What supply chain and food safety risks did Yum China Holdings, Inc. disclose in the risk factors of its 10-K? \\
\texttt{T1-3} & What cybersecurity and data privacy risks did ROKU, INC disclose in the risk factors of its 10-K? \\
\texttt{T1-4} & What cybersecurity and competition risks did SentinelOne, Inc. disclose in the risk factors of its 10-K? \\
\texttt{T1-5} & What regulatory and reimbursement risks did Privia Health Group, Inc. disclose in the risk factors of its 10-K? \\
\texttt{T2-1} & Which audit firm appears as the auditor for the most companies in this dataset, and how many distinct companies does it audit? \\
\texttt{T2-2} & Which company discloses the most subsidiaries in its EX-21 exhibit, and how many does it disclose? \\
\texttt{T2-3} & What is the most common jurisdiction of incorporation among all disclosed subsidiaries, and how many subsidiaries are incorporated there? \\
\texttt{T2-4} & Which SEC form type appears most often in this dataset, and how many filings of that type are there? \\
\texttt{T2-5} & Which issuer has the most distinct insiders filing Forms 3, 4 or 5 against it, and how many distinct insiders are there? \\
\texttt{T3-1} & Brian Wendling holds an insider role at more than one company in this dataset. For every company where they are an officer or director, list that company's disclosed subsidiaries and summ... \\
\texttt{T3-2} & Monica Lozano holds an insider role at more than one company in this dataset. For every company where they are an officer or director, list that company's disclosed subsidiaries and summa... \\
\texttt{T3-3} & William Giles holds an insider role at more than one company in this dataset. For every company where they are an officer or director, list that company's disclosed subsidiaries and summa... \\
\texttt{T3-4} & Judy Bruner holds an insider role at more than one company in this dataset. For every company where they are an officer or director, list that company's disclosed subsidiaries and summari... \\
\texttt{T3-5} & Gregg Sengstack holds an insider role at more than one company in this dataset. For every company where they are an officer or director, list that company's disclosed subsidiaries and sum... \\
\texttt{T4-1} & Is ''American Airlines, Inc.'' listed as a subsidiary in another company's EX-21 exhibit while also being an SEC filer in its own right? If so, name the parent that lists it and confirm w... \\
\texttt{T4-2} & Is ''AEP Texas Inc.'' listed as a subsidiary in another company's EX-21 exhibit while also being an SEC filer in its own right? If so, name the parent that lists it and confirm whether it... \\
\texttt{T4-3} & Is ''Athene Holding Ltd.'' listed as a subsidiary in another company's EX-21 exhibit while also being an SEC filer in its own right? If so, name the parent that lists it and confirm wheth... \\
\texttt{T4-4} & Is ''Alexander's, Inc.'' listed as a subsidiary in another company's EX-21 exhibit while also being an SEC filer in its own right? If so, name the parent that lists it and confirm whether... \\
\texttt{T4-5} & Is ''AllianceBernstein Holding L.P.'' listed as a subsidiary in another company's EX-21 exhibit while also being an SEC filer in its own right? If so, name the parent that lists it and co... \\
\bottomrule
\end{tabular}

\caption{The twenty benchmark questions. \texttt{T1} single-document semantic,
\texttt{T2} structured aggregation, \texttt{T3} multi-hop relational plus text,
\texttt{T4} cross-document entity resolution.}
\label{tab:questions}
\end{widetab}

\section{Prompt Templates and Schema Definitions}
\label{app:prompts}

This section reproduces the exact system prompts, schema documentation, and
deterministic query templates used across all three retrieval architectures,
verbatim from the implementation modules (\texttt{mode\_sql.py},
\texttt{mode\_vector.py}, and \texttt{mode\_graphrag.py}).

\subsection{Text-to-SQL Prompts and Schema}
\label{app:sql-prompts}

\paragraph{Relational Schema Description (\texttt{SCHEMA}):}
\begin{scriptsize}
\begin{verbatim}
-- PostgreSQL 16. Data slice: SEC EDGAR filings
-- from xbrlrss-2022-01 (Jan 2022) plus
-- Form 3/4/5 insider ownership dataset.

company(cik PK, name, sic, sic_description,
        fiscal_year_end, source)
  -- source='rss': company has filings/text;
  -- 'form345': ownership edges only.

filing(accession_number PK CHAR(20),
       company_cik FK, form_type,
       filing_date DATE, period_of_report DATE,
       index_url, primary_doc_url)

filing_document(accession_number FK, sequence,
                doc_type, filename, description,
                size_bytes, inline_xbrl, url)
  -- doc_type like '10-K', 'EX-21.1', 'EX-99.1'.

filing_section(accession_number FK, item_code,
               item_title, section_text TEXT,
               char_len, company_cik,
               filing_date)
  -- item_code IN ('1','1A','1B','2','3','7','7A').
  -- '1A' = Risk Factors. Full narrative text.
  -- Use ILIKE for keyword search.

subsidiary(accession_number FK, parent_cik,
           subsidiary_name, name_normalized,
           jurisdiction)
  -- parsed from EX-21. Subsidiaries have NO cik.

reporting_owner(accession_number, owner_cik,
                owner_name, issuer_cik,
                issuer_name, relationship,
                is_officer BOOL, is_director BOOL,
                is_ten_pct_owner BOOL,
                is_other BOOL, officer_title,
                filing_date, period_of_report)
  -- one row per (Form 3/4/5 filing, insider).
  -- owner_name uses EDGAR convention
  -- SURNAME FIRST, e.g. 'POLK DENNIS'.
  -- Match single surname: ILIKE '%polk%'.
  -- Use boolean columns over relationship.

filing_auditor(accession_number PK, company_cik,
               auditor_name, auditor_location,
               pcaob_firm_id, fiscal_year)

v_risk_factors(accession_number, company_cik,
               company_name, sic,
               sic_description, filing_date,
               char_len, section_text)
\end{verbatim}
\end{scriptsize}

\paragraph{Query Generation System Prompt (\texttt{GEN\_SYSTEM}):}
\begin{scriptsize}
\begin{verbatim}
You write a single read-only PostgreSQL query
answering the user's question.

Rules:
  - Output ONLY the SQL. No prose, no markdown.
  - SELECT or WITH only. Never write data.
  - Always LIMIT to at most 50 rows.
  - section_text columns are large (avg 75,000c):
    never SELECT raw. Note left(section_text, 400)
    returns only section HEADER. To pull passage
    around keyword, use:
      substring(section_text
        from greatest(1, position('...' in
                      lower(section_text)) - 400)
        for 1200)
    or:
      ts_headline('english', section_text,
                  plainto_tsquery('english', '...'),
                  'MaxFragments=3, MaxWords=60')
  - Prefer boolean flags (is_officer, is_director,
    is_ten_pct_owner) over parsing relationship.
  - Beware the LIMIT trap: if question covers
    several entities with hundreds of children,
    one entity consumes whole LIMIT. Aggregate
    per entity (count(*) plus string_agg).
  - If question asks whether ANY row satisfies
    condition, put condition in WHERE.
    Never infer negative from limited sample.
\end{verbatim}
\end{scriptsize}

\paragraph{Answer Synthesis System Prompt (\texttt{ANSWER\_SYSTEM}):}
\begin{scriptsize}
\begin{verbatim}
You are a financial-disclosure analyst.
Answer using ONLY the query result rows.
Cite accession numbers for factual claims.
If rows are empty or do not support answer,
say exactly that and name what is missing -
do not speculate. Concise: under 300 words.
\end{verbatim}
\end{scriptsize}

\subsection{Vector RAG Prompt}
\label{app:vec-prompts}

\paragraph{Answer Synthesis System Prompt (\texttt{SYSTEM}):}
\begin{scriptsize}
\begin{verbatim}
You are a financial-disclosure analyst.
Answer ONLY from the provided excerpts.
Cite accession number for every claim,
like [0001177394-22-000010]. If excerpts do not
contain answer, say exactly what is missing.
Never guess company relationship, ownership,
or executive role not stated verbatim.
Concise: under 300 words.
\end{verbatim}
\end{scriptsize}

\subsection{GraphRAG Prompts and Schema}
\label{app:graph-prompts}

\paragraph{Graph Schema Description (\texttt{GRAPH\_SCHEMA}):}
\begin{scriptsize}
\begin{verbatim}
Neo4j 5. Node labels and properties:
  (:Company   {cik INT, name, sic,
               sicDescription, source})
  (:Person    {cik INT, name})
  (:Subsidiary{nameNormalized, name,
               jurisdiction})
  (:AuditFirm {firmKey, name, pcaobFirmId})
  (:Filing    {accessionNumber, formType,
               filingDate DATE,
               periodOfReport DATE,
               hasRiskFactors BOOL,
               riskFactorChars INT, indexUrl})

Relationships (directions matter, follow exactly):
  (:Company)-[:FILED]->(:Filing)
  (:Person)-[:OFFICER_OF  {filingId, title,
                           filingDate}]->(:Company)
  (:Person)-[:DIRECTOR_OF {filingId,
                           filingDate}]->(:Company)
  (:Person)-[:OWNS_SHARES {filingId, basis,
                           filingDate}]->(:Company)
  (:Subsidiary)-[:SUBSIDIARY_OF {filingId,
                           jurisdiction}]->(:Company)
  (:Company)-[:AUDITED_BY {filingId,
                           fiscalYear}]->(:AuditFirm)
  (:Subsidiary)-[:RESOLVES_TO]->(:Company)

Notes:
  - Person names stored EDGAR-style, surname first:
    'POLK DENNIS'. Match:
    toLower(p.name) CONTAINS 'polk'.
  - Only :Filing nodes with hasRiskFactors=true
    have Item 1A in Postgres.
  - Always RETURN filingId of edges and
    accessionNumber of any :Filing touched.
\end{verbatim}
\end{scriptsize}

\paragraph{Planning System Prompt (\texttt{PLAN\_SYSTEM}):}
\begin{scriptsize}
\begin{verbatim}
You plan a GraphRAG retrieval over an SEC graph.
Return ONLY JSON object, no markdown:
  {"cypher": "<one read-only Cypher query>",
   "text_terms": ["<keyword>", ...]}

Cypher rules (stage 1 only):
  - Stage 1: identify WHICH entities matter.
    Return ONE ROW PER ENTITY with identifier
    (companyCik). Do NOT traverse to children;
    system expands entities in stage 2.
    Fan-out query wastes LIMIT on one entity.
  - Read-only. No CREATE/MERGE/SET/DELETE.
  - LIMIT 25 or fewer.
  - If question asks whether ANY node satisfies
    condition, put condition inside MATCH.
    Never scan whole label.
  - RETURN accessionNumber of every :Filing,
    aliased accessionNumber (or filingId),
    plus entity names.
  - filingId identifies WHICH DOCUMENT edge
    came from. NEVER join one edge's filingId
    to another's.
  - Do NOT collect() unbounded list into one row.
text_terms: 2-4 short keyword phrases to locate
relevant passage in risk-factor text (lowercase).
\end{verbatim}
\end{scriptsize}

\paragraph{Deterministic Expansion Query (\texttt{EXPAND\_CQL}):}
\begin{scriptsize}
\begin{verbatim}
UNWIND $ciks AS wanted
MATCH (c:Company {cik: wanted})
OPTIONAL MATCH (s:Subsidiary)-[so:SUBSIDIARY_OF]->(c)
WITH c, count(DISTINCT s) AS subsidiaryCount,
     collect(DISTINCT s.name)[..12]
       AS subsidiarySample,
     collect(DISTINCT so.filingId)[..1]
       AS ex21FilingId
OPTIONAL MATCH (c)-[:FILED]->(f:Filing)
  WHERE f.hasRiskFactors
WITH c, subsidiaryCount, subsidiarySample,
     ex21FilingId,
     collect(DISTINCT f.accessionNumber)
       AS riskFilingAccessions
OPTIONAL MATCH (p:Person)-[:OFFICER_OF|DIRECTOR_OF]->(c)
WITH c, subsidiaryCount, subsidiarySample,
     ex21FilingId, riskFilingAccessions,
     collect(DISTINCT p.name)[..6] AS someInsiders
OPTIONAL MATCH (self:Subsidiary)-[:RESOLVES_TO]->(c)
OPTIONAL MATCH (self)-[selfEdge:SUBSIDIARY_OF]->
               (parentCo:Company)
RETURN c.cik AS companyCik,
       c.name AS companyName,
       subsidiaryCount,
       subsidiarySample, ex21FilingId,
       riskFilingAccessions, someInsiders,
       collect(DISTINCT parentCo.name)
         AS alsoNamedAsSubsidiaryOf,
       collect(DISTINCT selfEdge.filingId)
         AS namedAsSubsidiaryInFilings
\end{verbatim}
\end{scriptsize}

\paragraph{Grounded Answer Synthesis System Prompt (\texttt{ANSWER\_SYSTEM}):}
\begin{scriptsize}
\begin{verbatim}
You are a financial-disclosure analyst. You are given
(a) structured graph facts, and (b) verbatim excerpts
from specific filings those facts point to. Answer
using only these. Complete per-company expansion block
is authoritative and NOT truncated. Traversal rows may
be truncated by LIMIT - never infer "none exist" from
them. Cite accession number for every claim. Name
traversal path in one line. If a hop produced nothing,
say so plainly. Concise: under 300 words. Name every
company found. For long subsidiary lists, give count
and 5 examples.
\end{verbatim}
\end{scriptsize}

\end{document}